\documentclass[a4paper,11pt]{article}
\pdfoutput=1 

\usepackage{jcappub} 

\usepackage{appendix}

\title{Neutrinos in the holographic dark energy model: constraints from latest measurements of expansion history and growth of structure}

\author[a]{Jing-Fei Zhang,}
\author[a]{Ming-Ming Zhao,}
\author[a]{Yun-He Li,}
\author[a,b,1]{Xin Zhang\note{Corresponding author.}}

\affiliation[a]{Department of Physics, College of Sciences, Northeastern University, \\Shenyang
110004, China}
\affiliation[b]{Center for High Energy Physics, Peking University, \\Beijing 100080, China}

\emailAdd{jfzhang@mail.neu.edu.cn}
\emailAdd{zhaomingmingsp@163.com}
\emailAdd{liyh19881206@126.com}
\emailAdd{zhangxin@mail.neu.edu.cn}

\abstract{The model of holographic dark energy (HDE) with massive neutrinos and/or dark radiation is investigated in detail. The background and perturbation evolutions in the HDE model are calculated. We employ the PPF approach to overcome the gravity instability difficulty (perturbation divergence of dark energy) led by the equation-of-state parameter $w$ evolving across the phantom divide $w=-1$ in the HDE model with $c<1$. We thus derive the evolutions of density perturbations of various components and metric fluctuations in the HDE model. The impacts of massive neutrino and dark radiation on the CMB anisotropy power spectrum and the matter power spectrum in the HDE scenario are discussed. Furthermore, we constrain the models of HDE with massive neutrinos and/or dark radiation by using the latest measurements of expansion history and growth of structure, including the Planck CMB temperature data, the baryon acoustic oscillation data, the JLA supernova data, the Hubble constant direct measurement, the cosmic shear data of weak lensing, the Planck CMB lensing data, and the redshift space distortions data.
We find that $\sum m_\nu<0.186$ eV (95\% CL) and $N_{\rm eff}=3.75^{+0.28}_{-0.32}$ in the HDE model from the constraints of these data.}

\begin{document}
\maketitle
\flushbottom

\section{Introduction}
\label{intro}

Detection of neutrino oscillations has indicated that neutrinos have masses. However, the neutrino oscillation experiments cannot measure the absolute masses of neutrinos, but can only measure the squared mass differences between the neutrino mass eigenstates. For example, the solar and reactor experiments observed $\Delta m_{21}^2\simeq 8\times 10^{-5}$
eV$^2$, and the atmospheric and accelerator beam experiments observed $\Delta m_{32}^2\simeq 3\times 10^{-3}$
eV$^2$. As a complementary to laboratory experiments, current cosmological data have been used to get information on the absolute scale of neutrino masses.
Actually, the current available cosmological data have been providing tight limits on the total mass of neutrinos.
See Refs. \cite{neu1,neu2,neu3} for reviews of neutrino cosmology.

The cosmic microwave background (CMB) observations have been used to constrain the neutrino mass and
the extra relativistic degrees of freedom (sometimes referred to as ``dark radiation'')~\cite{wmap5,wmap7,wmap9}.
In the base 6-parameter $\Lambda$CDM model, a normal mass hierarchy with $\sum m_\nu\approx 0.06$ eV (dominated by the heaviest neutrino mass eigenstate) is assumed.
When the base $\Lambda$CDM model is extended to allow for larger neutrino masses, one often assumes three species of degenerate massive neutrinos, neglecting the small differences between mass eigenstates. The Planck CMB temperature power spectrum provided the tight limits on
the total mass of active neutrinos, $\sum m_\nu$, and the  effective number of relativistic species, $N_{\rm eff}$~\cite{planck}.
For example, the Planck+WP+highL data combination (here, WP denotes the WMAP 9-year polarization data, and
highL denotes the ACT and SPT temperature data) gives the 95\% confidence level (CL) limits:
$\sum m_\nu<0.66$ eV for the case of no extra relics ($N_{\rm eff}=3.046$) and $N_{\rm eff}=3.36^{+0.68}_{-0.64}$ for
the case of minimal-mass normal hierarchy for the neutrino masses (only one massive eigenstate with $m_\nu=0.06$ eV). 
Note here that, unless otherwise specified, Planck data in this paper refer to the Planck 2013 release. 
Late-time geometric measurements can be used to help reduce some geometric degeneracies and thus improve constraints.
Therefore, the baryon acoustic oscillation (BAO) data are very useful in the parameter estimation.
The Planck+WP+highL+BAO data combination changes the above limits to:
$\sum m_\nu<0.23$ eV for the case of no extra relics and $N_{\rm eff}=3.30^{+0.54}_{-0.51}$ for the case of minimal-mass
normal hierarchy model.\footnote{During the completion of this paper, the Planck 2015 results appeared on arXiv \cite{Planck:2015xua} (but the new data have not been released currently). According to the latest Planck 2015 data,
it is found that $\sum m_\nu<0.72$ eV from Planck TT+lowP and $\sum m_\nu<0.21$ eV from Planck TT+lowP+BAO;
$N_{\rm eff}=3.13\pm 0.32$ from Planck TT+lowP and $N_{\rm eff}=3.15\pm 0.23$ from Planck TT+lowP+BAO.
Here ``lowP'' denotes the Planck low-$\ell$ temperature-polarization data.}

Since the Planck data are in tension with the weak lensing measurements and the abundance of rich clusters, recently there has been significant interest in larger neutrino masses
because neutrino free streaming provides a possible way to lower the late-time fluctuation amplitude $\sigma_8$ and thereby could reconcile the above tensions.
For the relevant discussions, see, e.g., Refs. \cite{snu1,snu2,snu3,zx14,WHu14,sterile2,sterile3,nus14a,nus14b}. But it was also argued by some authors \cite{Leistedt:2014sia} that the larger neutrino masses offer only a marginal improvement compared to the base $\Lambda$CDM model. Anyway, the use of the large-scale structure measurements could help constrain the neutrino mass significantly.

The cases of neutrinos and dark radiation in a dynamical dark energy model have been discussed \cite{zxjcap14}. In Ref. \cite{zxjcap14}, the $\Lambda$CDM+neutrino/dark radiation models are extended by replacing the cosmological constant with the dynamical dark energy with constant $w$. The corresponding $w$CDM-based models are also constrained by using the recent observational data in Ref. \cite{zxjcap14}.
In this paper, we will further extend the relevant discussion. We will consider the cases of neutrinos and/or dark radiation in the holographic dark energy model.

The holographic dark energy (HDE) model \cite{li04} originates from the holographic principle of quantum gravity, and so it remains significant interest in cosmology.
It is expected that theoretical and phenomenological studies on HDE might provide important clues for the bottom-up exploration of a full quantum theory of gravity.
There have been numerous studies on the theoretical implications and observational constraints of the HDE model \cite{hde1,hde2,hde3,hde4,hde5,hde6,hde7,hde8,hde9,hde10,hde11,hde12,hde13,hde14,hde15,hde16,hde17,hde18,hde19,hde20,hde21,hde22}.
In particular, the neutrino mass in the HDE model has been constrained in the light of the WMAP 7-year data in Ref. \cite{hde19}.
In this paper, we will make a more sophisticated analysis; we will discuss the constraints on the neutrino mass and/or the extra relativistic degrees of freedom in the HDE model by using the Planck CMB data plus other measurements of expansion history and growth of structure.

We will consider three models in this paper, i.e., (i) the HDE+$\sum m_\nu$ model, (ii) the HDE+$N_{\rm eff}$ model, and (iii) the HDE+$\sum m_\nu$+$N_{\rm eff}$ model.
We will constrain the models by using the latest observational data. The basic data combination we use is the CMB+BAO.
Furthermore, we combine other geometric measurements, i.e., the type Ia supernovae (SN) JLA data and the direct determination of the Hubble constant $H_0$.
Finally, we also consider to further combine the measurements of growth of structure, i.e., the weak lensing (WL) data and the redshift space distortions (RSD) data.

The paper is organized as follows. In Sec. \ref{sec:hde}, we briefly describe the model of holographic dark energy with massive neutrinos and dark radiation.
For the cases of $c<1$ (note that here $c$ is not the speed of light, but the parameter of HDE) in the HDE model, the dark-energy equation-of-state parameter (EOS) $w$ crosses the phantom divide $w=-1$. Usually, the perturbation instability appears when $w$ crosses $-1$. In order to overcome this difficulty, we employ the ``parametrized post-Friedmann'' (PPF) approach \cite{PPF} to treat the perturbations in dark energy. We will illustrate the density perturbations of various components and the metric fluctuations in the HDE model.
Also, we will discuss the impacts of massive neutrinos and dark radiation on the CMB anisotropy spectrum and the matter power spectrum.
In Sec. \ref{sec:obs}, we use the latest observational data to constrain the models. We will focus on the constraint results of $c$, $\Omega_m$, $H_0$, $\sigma_8$, $\sum m_\nu$, and $N_{\rm eff}$.
Conclusion is given in Sec. \ref{sec:concl}.

\section{A brief description for the model of holographic dark energy with massive neutrinos and dark radiation}\label{sec:hde}

The cosmological constant suffers from the severe theoretical challenge, i.e., it cannot be understood why the theoretical value of $\Lambda$ is greater than the observational value by many orders of magnitude. The cosmological constant is equivalent to the vacuum energy density, and thus its value is determined by the sum of the zero-point energy of each mode of all the quantum fields. Thus we have the vacuum energy density $\rho_\Lambda\simeq k_{\rm max}^4/(16\pi^2)$, where $k_{\rm max}$ is the imposed momentum ultraviolet (UV) cutoff. When taking the UV cutoff to be the Planck scale ($10^{19}$ GeV), where the quantum field theory in a classical spacetime metric is expected to breakdown, the vacuum energy density would exceed the critical density by some 120 orders of magnitude.

Obviously, one should not calculate the value of $\Lambda$ in the context without gravity. In the absence of a full theory of quantum gravity, one has to explore the potential solution of the cosmological constant problem through combining the holographic principle of quantum gravity with the effective theory of quantum fields. That is the origination of the HDE model.

Considering gravity in a quantum field system, the conventional
local quantum field theory would break down due to the too many
degrees of freedom that would cause the formation of a black hole.
However, once the holographic principle is considered, the number of
degrees of freedom can be reduced.

One could put an energy bound
on the vacuum energy density, $\rho_\Lambda L^3\leq M_{\rm Pl}^2 L$,
where $M_{\rm Pl}$ is the reduced Planck mass, which implies that
the total energy in a spatial region with size $L$ should not exceed
the mass of a black hole with the same size~\cite{Cohen:1998zx}. The
largest length size compatible with this bound is the infrared (IR)
cutoff size of this effective quantum field theory. Evidently, the
holographic principle gives rise to a dark energy model based on
the effective quantum field theory with a UV/IR duality. From the
UV/IR correspondence, the UV problem of dark energy can be converted
into an IR problem. A given IR scale can saturate that bound, and thus
one can write the dark energy density as \cite{li04}
\begin{equation}
\rho_\Lambda=3c^2M_{\rm
Pl}^2L^{-2},
\end{equation}
where $c$ is a dimensionless parameter
characterizing all of the uncertainties of the
theory. This indicates that the UV cutoff of the theory would not be
fixed but run with the evolution of the IR cutoff, i.e., $k_{\rm
max}\propto L^{-1/2}$.

The HDE model
chooses the event horizon of the universe
\begin{equation}
R_{\rm EH}(t)=a(t)\int_t^\infty{dt'\over a(t')}
\end{equation}
as the IR cutoff $L$ of the
theory, explaining the fine-tuning problem and the coincidence
problem at the same time in some degree~\cite{li04}. Actually, it is
clear to see that the holographic dark energy is essentially a
holographic vacuum energy. However, this
holographic vacuum energy does not behave like a usual vacuum
energy, owing to the fact that its equation of state parameter
$w$ is not equal to $-1$.

The HDE is a dynamical dark energy. The parameter $c$ plays a crucial role in determining the evolution of the HDE. At the early times ($t\to 0$ or $z\to \infty$), the EOS of HDE $w\to -1/3$. The value of $c$ determines the subsequent evolution of $w$. In the infinite future ($t\to \infty$ or $z\to -1$), we have $w\to -1/3-2/(3c)$. Thus, when $c<1$, we find that $w$ of HDE crosses $-1$ during the cosmological evolution.

\subsection{Background and perturbation evolutions of holographic dark energy}

Consider a spatially flat universe with HDE, matter, and radiation. The Friedmann equation reads
\begin{equation}
3M_{\rm Pl}^2H^2=\rho_\Lambda+\rho_c+\rho_b+\rho_\gamma+\rho_\nu,
\end{equation}
where $\rho_\Lambda$, $\rho_c$, $\rho_b$, $\rho_\gamma$, and $\rho_\nu$ stand for the energy densities of HDE, cold dark matter (CDM), baryons, photons, and neutrinos.
The CDM and baryons are non-relativistic matter, and the photons are of course radiation. The neutrinos contribute to the radiation density at the early times, but they behave as matter after the non-relativistic transition. Actually, there may be some extra radiation, usually referred to as ``dark radiation''. Their contribution to the total radiation content can be parametrized in terms of the effective number of neutrinos, $N_{\rm eff}$. So the energy density of total radiation can be given by the following relation
\begin{equation}
\rho_r=\left[1+{7\over 8}\left({4\over 11}\right)^{4/3}N_{\rm eff}\right]\rho_\gamma,
\end{equation}
where we see that $\rho_r$ has been normalized by the energy density of photons due to the fact that its present-day value is measured from the CMB temperature.

Since the HDE model has been investigated extensively, here we directly give the background evolution equations that are coupled differential equations governing the functions $\Omega_\Lambda(z)$ and $E(z)$. The equations are \cite{hde19,hde20}:
\begin{equation}\label{deq1}
{1\over E(z)}{dE(z)\over dz}=-{\Omega_\Lambda(z)\over 1+z}\left({1\over c}\sqrt{\Omega_\Lambda(z)}+{\Omega_\Lambda(z)-\Omega_r(z)-3\over 2\Omega_\Lambda(z)}\right),
\end{equation}
\begin{equation}\label{deq2}
{d\Omega_\Lambda(z)\over dz}=-{2\Omega_\Lambda(z)(1-\Omega_\Lambda(z))\over 1+z}\left({1\over c}\sqrt{\Omega_\Lambda(z)}+{1-\Omega_\Lambda(z)+\Omega_r(z)\over 2(1-\Omega_\Lambda(z))}\right),
\end{equation}
where $E(z)\equiv H(z)/H_0$ is the dimensionless Hubble expansion rate, and $\Omega_r(z)=\Omega_r(1+z)^4/E(z)^2$. The initial condition for solving the above differential equations are:
$E(0)=1$ and $\Omega_\Lambda(0)=1-\Omega_m-\Omega_r$.

The EOS of HDE is given by the following relation \cite{li04}
\begin{equation}
w(z)={1\over 3}-{2\over 3c}\sqrt{\Omega_\Lambda(z)}.
\end{equation}
Thus, once the function $\Omega_\Lambda(z)$ is derived as a solution to the differential equations (\ref{deq1}) and (\ref{deq2}), the evolution of $w(z)$ is also entirely determined.

The linear metric and matter density perturbations of the HDE model can be calculated by using the formalism of Ma and Bertschinger \cite{Ma:1995ey}. But when treating the dark energy perturbations, one should be very careful about the divergence problem at the $w=-1$ crossing. We follow the WMAP and Planck collaborations to deal with this issue by using the PPF approach \cite{PPF}.

The perturbations of dark energy can be described by four variables: density fluctuation $\delta\rho_{\rm de}$, velocity $v_{\rm de}$, pressure fluctuation $\delta p_{\rm de}$, and anisotropic stress $\Pi_{\rm de}$. For these quantities, the evolution equations of $\delta\rho_{\rm de}$ and $v_{\rm de}$ are determined by continuity and Navier-Stokes equations, and $\Pi_{\rm de}$ vanishes for the linear perturbations. In usual fluid approach, to complete the system, one need to specify the relationship between $\delta p_{\rm de}$ and $\delta\rho_{\rm de}$ by defining the rest-frame sound speed for dark energy. But the PPF method abandons this condition. The PPF approach replaces the condition of pressure perturbation with a direct relationship between the momentum densities of dark energy and other components on large scales, which determines the velocity $v_{\rm de}$. Actually, once $v_{\rm de}$ is determined, $\delta p_{\rm de}$ follows by the momentum conservation, and in this method no divergence will appear when $w$ crosses the phantom divide.
The details about the PPF description in the HDE model can be found in Sec. 2.2 of Ref. \cite{hde19}. For more details for the PPF code, we refer the reader to Ref. \cite{PPF}. Note also that the PPF framework for interacting dark energy has been established recently \cite{ppfide1,ppfide2}.

\begin{figure*}
\includegraphics[width=7.4cm]{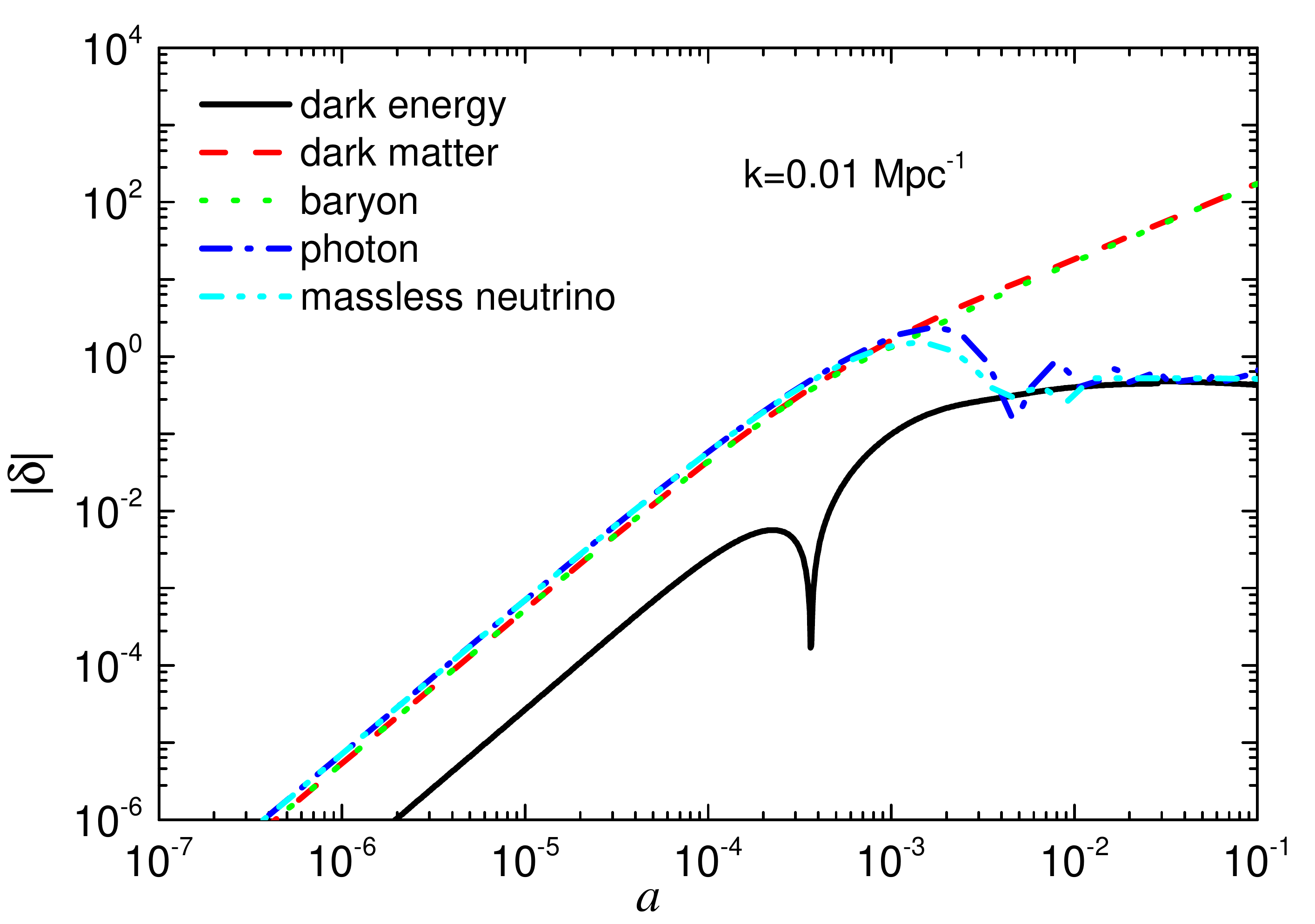}
\includegraphics[width=7.4cm]{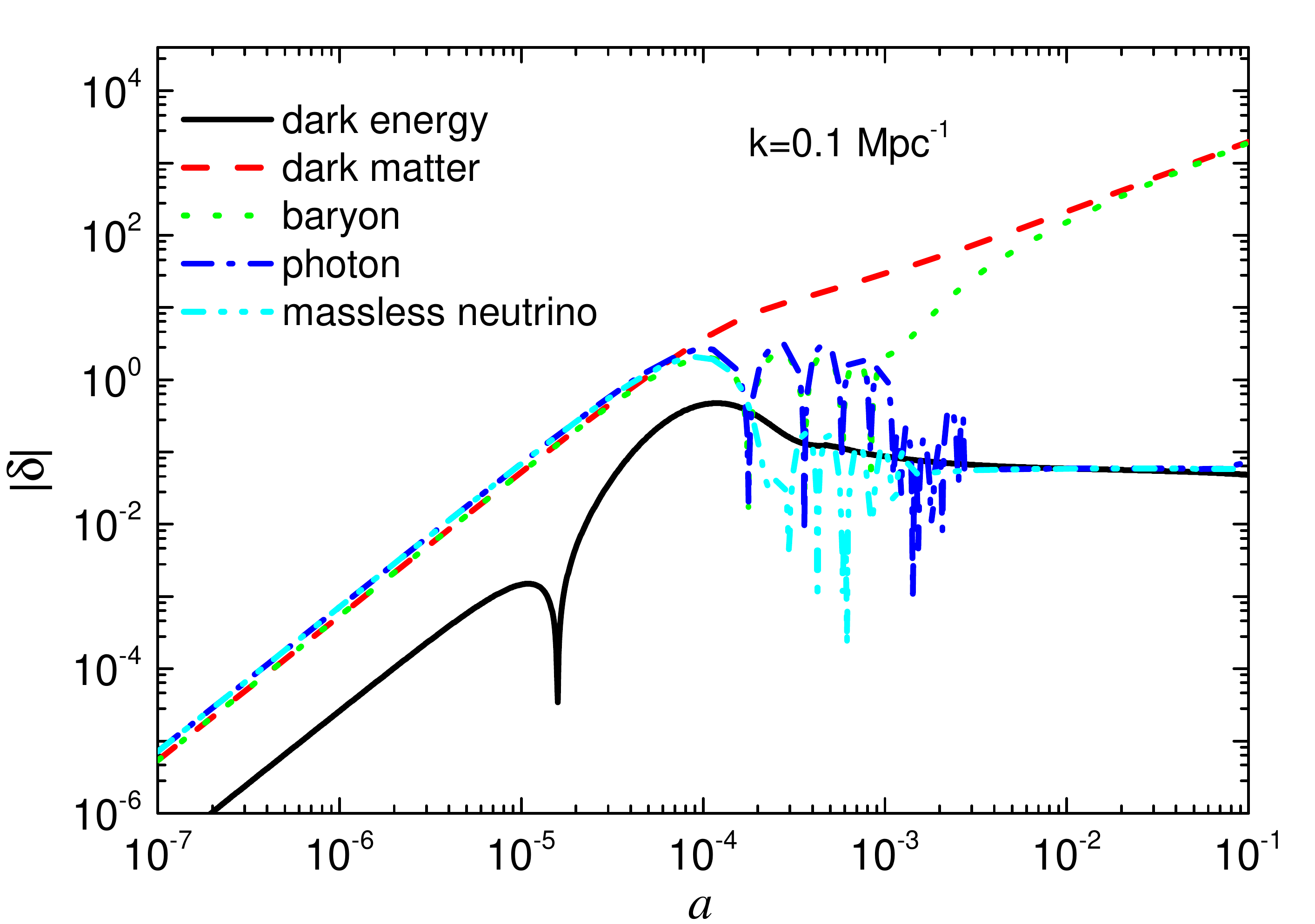}\\
\includegraphics[width=7.4cm]{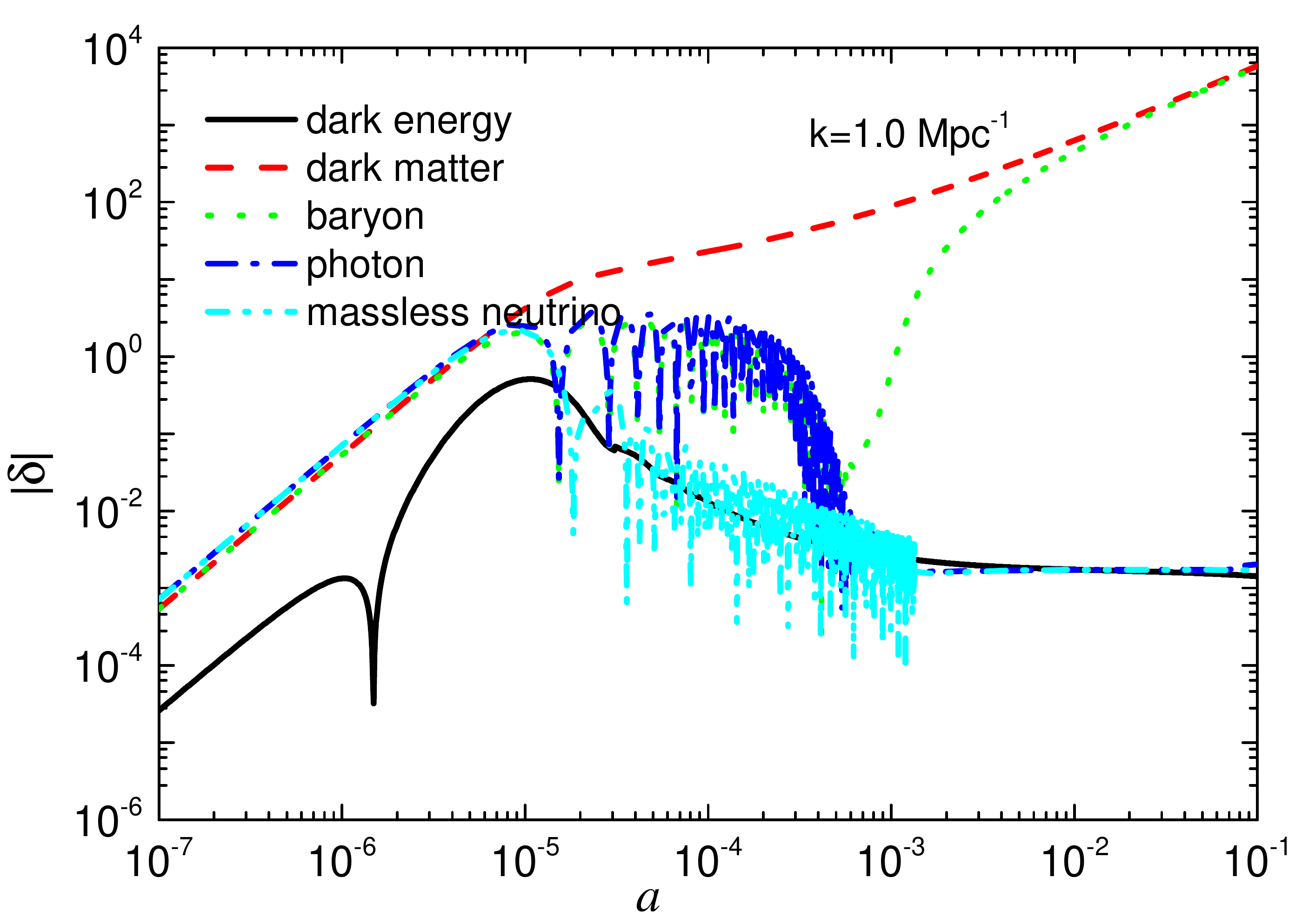}
\includegraphics[width=7.5cm]{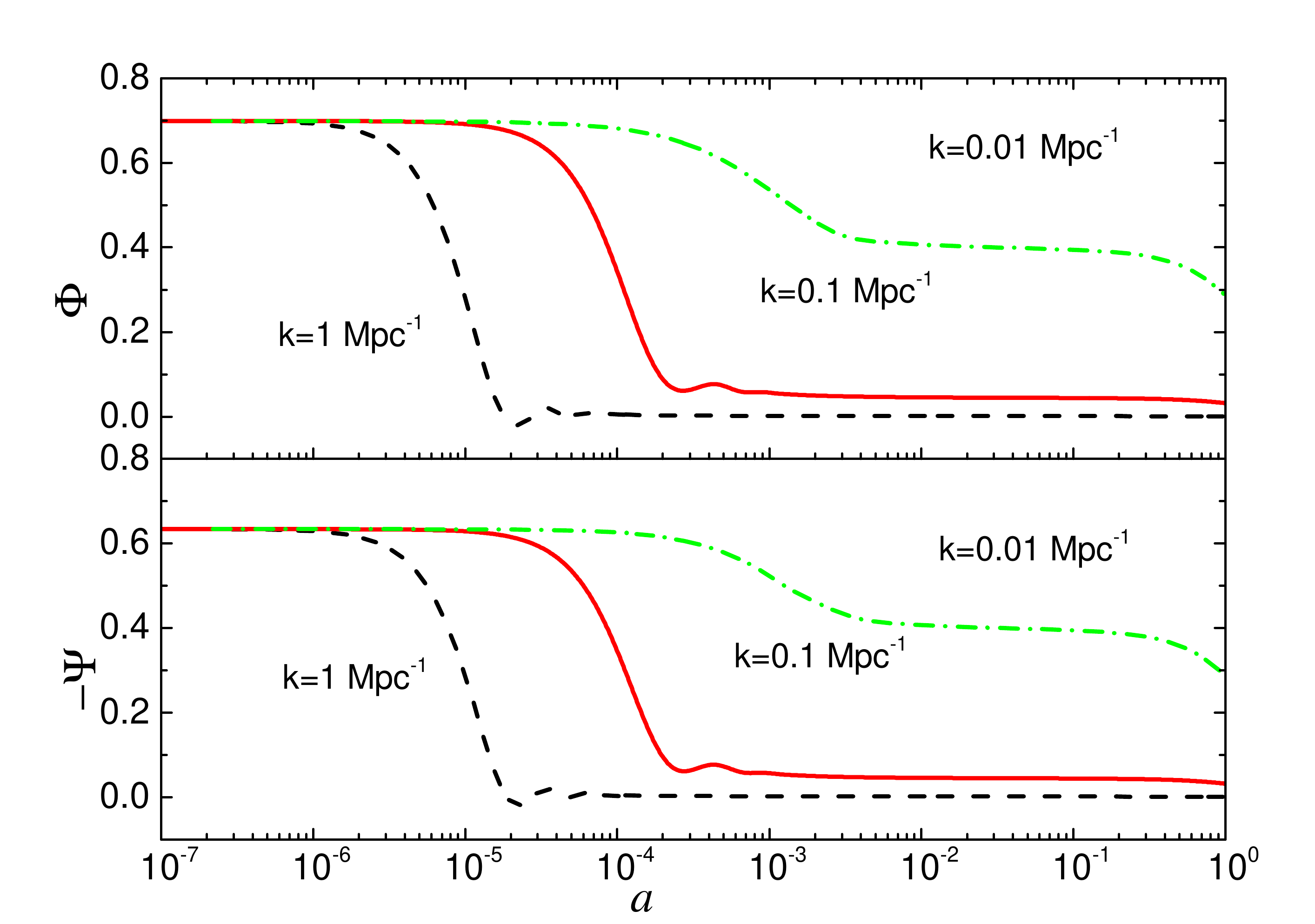}
\caption{\label{fig1} The evolutions of matter and metric perturbations in the HDE model at $k $= 0.01$\rm\, Mpc^{-1}$, 0.1$ \rm \, Mpc^{-1}$, and 1.0$ \rm\, Mpc^{-1}$.
Here the matter density perturbations are calculated in the synchronous gauge, and the metric perturbations $\Phi$ and $\Psi$ are the gauge-invariant variables. We fix $\sum m_\nu=0.06$ eV and $N_{\rm eff}=3.046$, and other parameters are fixed to be the best-fit values from Planck.}
\end{figure*}

Figure \ref{fig1} shows a concrete example of the numerical calculation for the evolution of matter density perturbations and metric fluctuations in the HDE model with $c=0.8$. To make a comparison for different scales, we choose three scales, i.e., $k=0.01$, 0.1, and 1.0 Mpc$^{-1}$, to illustrate.
In this example, other parameters are fixed as the values in the 6-parameter $\Lambda$CDM model fitting to the Planck data. In particular, we note that the normal hierarchy minimal mass assumption is made for the neutrino masses, i.e., $\sum m_\nu=0.06$ eV, and $N_{\rm eff}$ is fixed to be 3.046. We show the evolution of the density perturbations of HDE, CDM, baryons, photons, and massless neutrinos in the synchronous gauge. Outside the horizon, the behavior of the density perturbations is strongly gauge-dependent. In the synchronous gauge, before the horizon crossing, all the density contrasts grow; after the horizon crossing, they come into causal contact and become nearly independent of the coordinate choices. For the HDE, we find that after the horizon crossing, the density perturbations in HDE become constant on large scales (e.g., $k=0.01$ Mpc$^{-1}$) and decrease quickly on small scales (e.g., $k=0.1$ and 1.0 Mpc$^{-1}$).
Hence, the HDE would not cluster significantly on the sub-horizon size.
For all the cases, density perturbation of HDE is always smaller than that of CDM by several orders of magnitude, and is nearly in the same order of magnitude with that of radiation, and so the HDE perturbation almost does not affect the evolution of the CDM perturbation. The evolutions of the metric fluctuations $\Phi$ and $\Psi$ are also shown in this figure.

\subsection{Impacts of massive neutrinos and dark radiation on CMB anisotropy spectrum and matter power spectrum}

\begin{figure*}
\includegraphics[width=8cm]{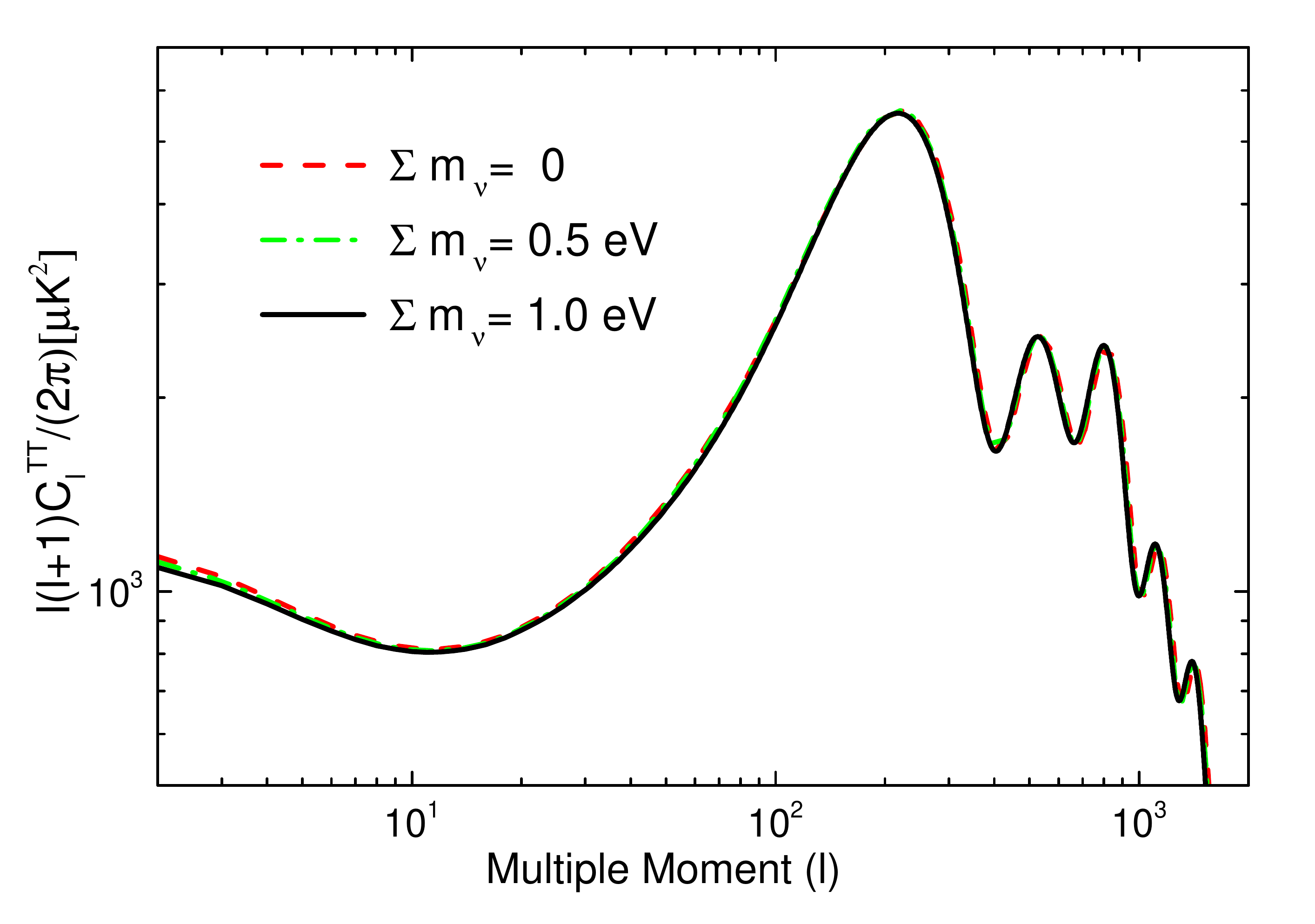}
\includegraphics[width=8cm]{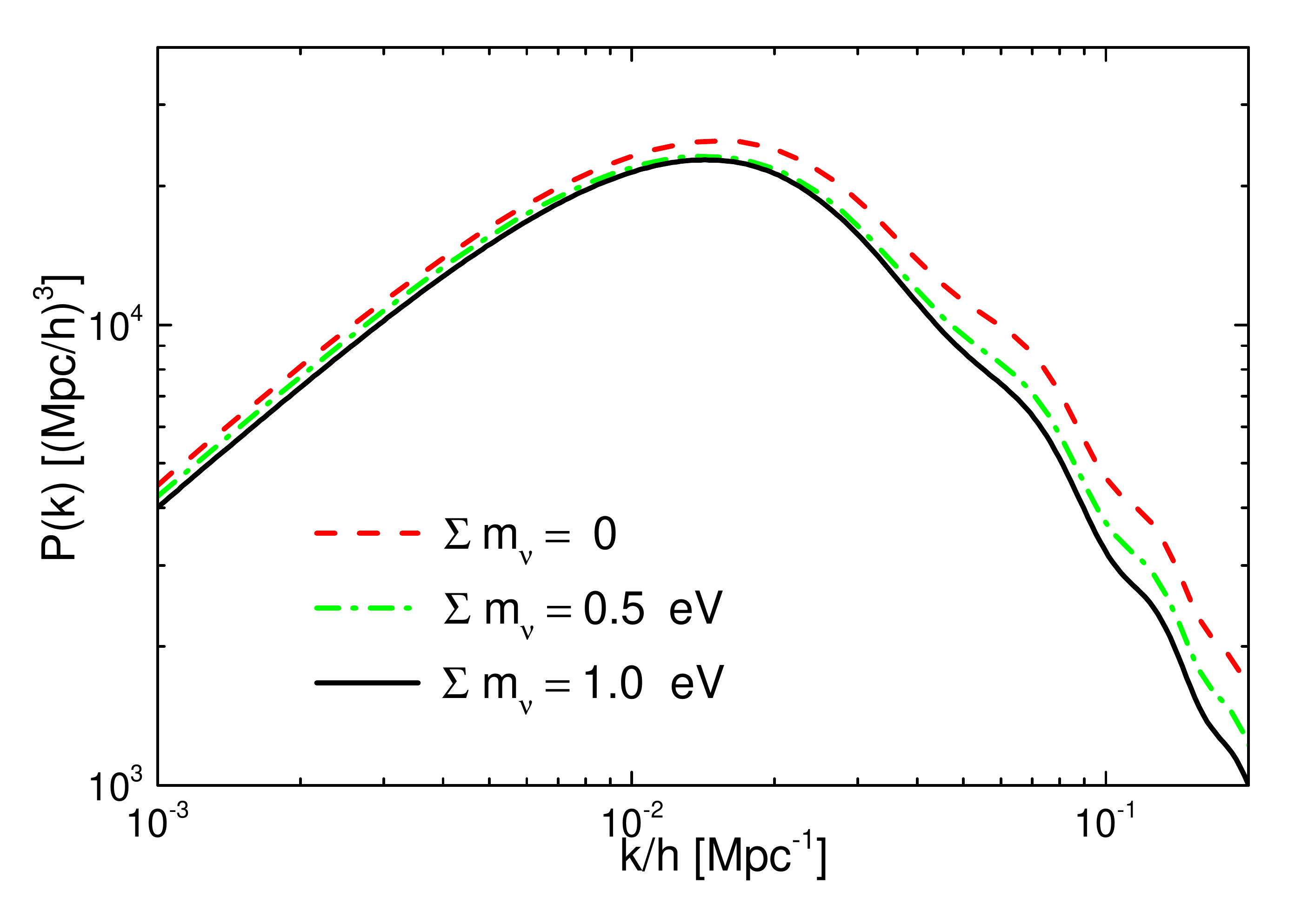}\\
\includegraphics[width=8cm]{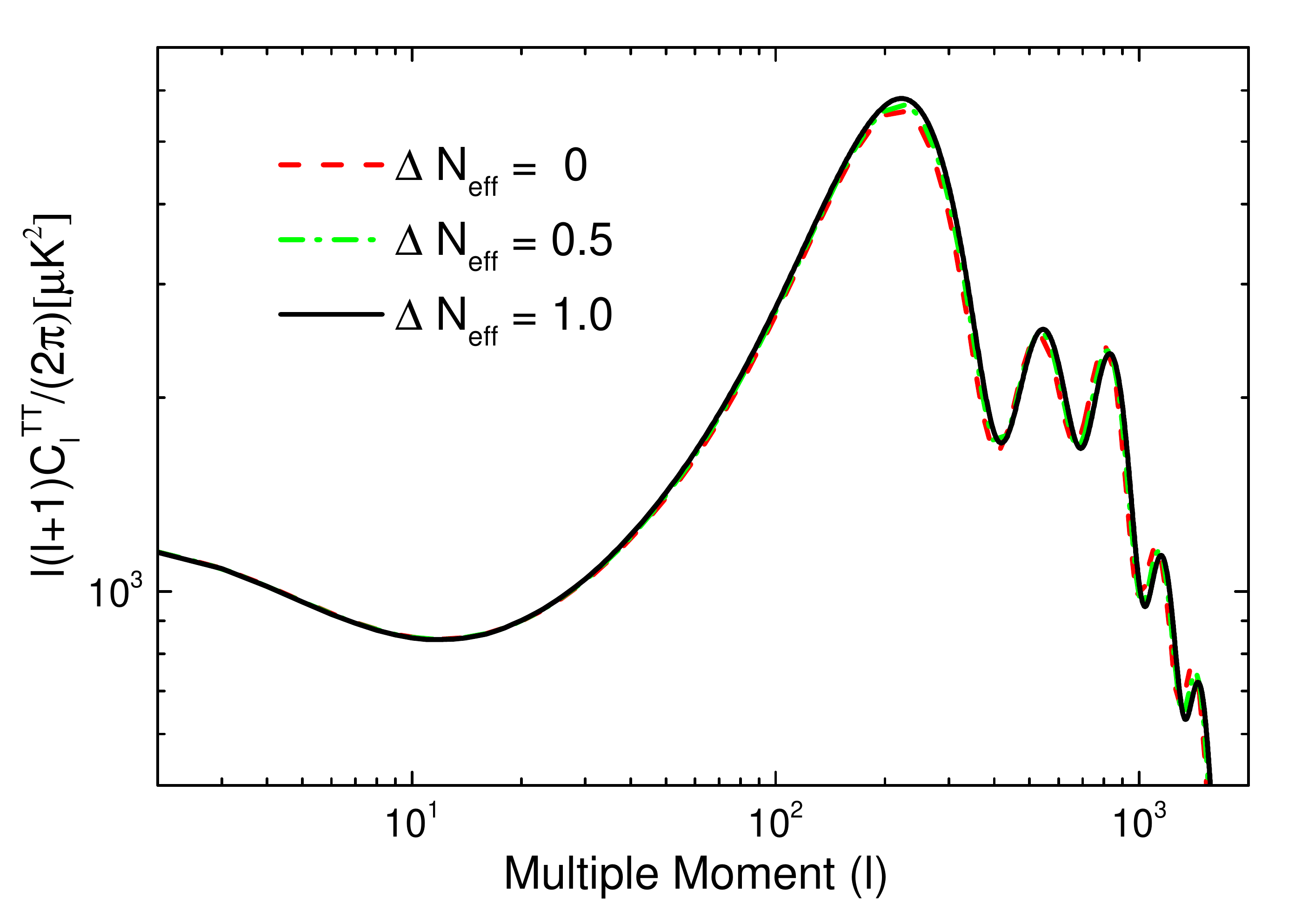}
\includegraphics[width=8cm]{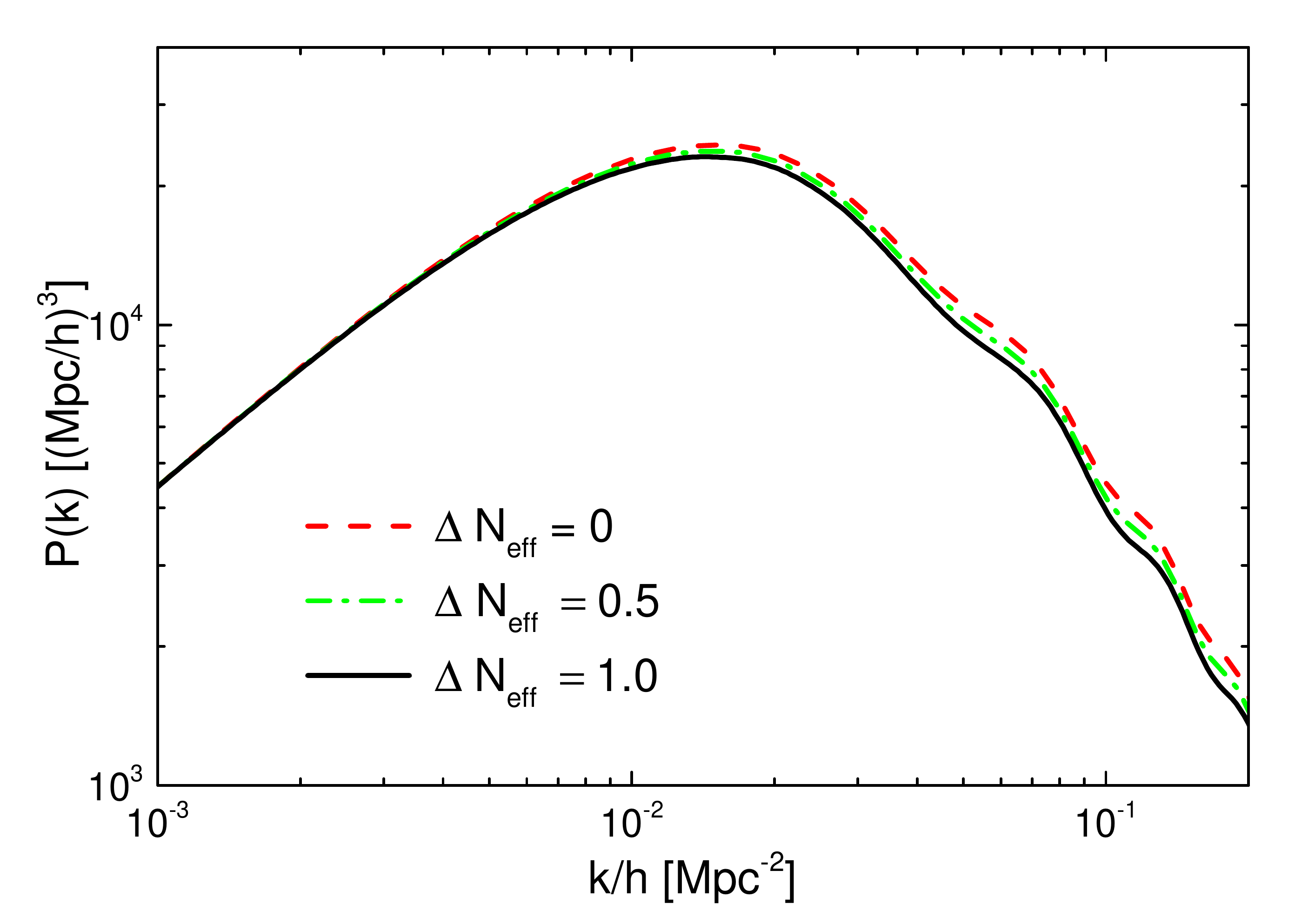}
\caption{\label{fig2}The CMB anisotropy power spectrum $C_\ell^{TT}$ and the matter power spectrum $P(k)$ in the HDE model with $c=0.8$.
In the upper panels, the parameter $\sum m_\nu$ is varied and other parameters are fixed; we choose $\sum m_\nu=0$, 0.5 eV, and 1.0 eV.
In the lower panels, the parameter $N_{\rm eff}$ is varied and other parameters are fixed; we choose $\Delta N_{\rm eff}=0$, 0.5, and 1.0.}
\end{figure*}

Neutrinos do not interact much at all with baryons for $z\ll 10^{10}$ (after thermal decoupling). As a result, neutrinos are treated as free-streaming particles. Since dark radiation behaves like massless neutrinos, it is also treated as free-streaming. They can affect the CMB anisotropy spectrum and matter power spectrum, thus providing a potential way to constrain them through CMB and large-scale structure (LSS) observations.

Massive neutrinos (with the total mass in the range from $10^{-3}$ eV to 1 eV) should be viewed as radiation at the time of equality and as non-relativistic matter today.
The change in the redshift of equality $z_{\rm eq}$ affects the position and amplitude of the peaks. The change in the non-relativistic matter density at late times leads to the two effects: impact the angular diameter distance to the last scattering surface, $D_A(z_\ast)$, that controls the overall position of CMB anisotropy spectrum features, and affect the slope of the low-$\ell$ tail of the CMB spectrum through the late Integrated Sachs-Wolfe (ISW) effect.

Neutrinos with masses can affect the matter power spectrum. On large scales ($k<k_{\rm nr}$, with $k_{\rm nr}\simeq 0.018\Omega_m^{1/2}({m_\nu\over1 {\rm eV}})^{1/2}~h$ Mpc$^{-1}$),
the neutrino free-streaming can be ignored, and so the neutrino perturbations are nearly indistinguishable from the CDM ones. Thus, the neutrino mass almost does not impact on the large-scale power spectrum. But the small-scale matter power spectrum $P(k)$ with $k>k_{\rm nr}$ is reduced by the massive neutrinos. Due to the free-streaming of neutrinos, massive neutrinos do not cluster on small scales. Also, the growth rate of CDM perturbations is reduced via the absence of gravitational back-reaction effects of free-streaming neutrinos.

The dark radiation (parametrized by $\Delta N_{\rm eff}=N_{\rm eff}-3.046$) affects the CMB power spectrum in several ways. First, the extra radiation density increases the early expansion rate, so the angular size of the acoustic scale, $\theta_\ast\equiv r_s/D_A$, is reduced, which determines the peak positions. Second, the extra dark radiation density delays the time of equality, and so enhances the first and second peaks due to the early ISW effect. Third, as a free-streaming fluid, the dark radiation has a non-negligible anisotropic stress, and this changes the metric fluctuations during the radiation era and thus the temperature fluctuations on scales $\ell\gtrsim 130$, because these scales enter the horizon during the radiation-dominated era. Finally, the extra radiation density increases the expansion rate and thus increases the diffusion length relative to the sound horizon, which enhances the Silk damping of the small-scale anisotropy. The dark radiation affects the matter power spectrum is also through the free-streaming, thus only the small-scale powers are suppressed by larger $N_{\rm eff}$.

Figure \ref{fig2} shows an example of how massive neutrinos and dark radiation impact on the CMB anisotropy power spectrum $C_\ell^{TT}$ and the matter power spectrum $P(k)$ in the HDE model with $c=0.8$. In the upper panels, we show the cases of varying the neutrino mass. We choose $\sum m_\nu=0$, 0.5, and 1.0 eV as examples; at the same time, other parameters are fixed (including $N_{\rm eff}=3.046$).
We find that for the CMB power spectrum the only observed differences are for $2<\ell<50$ due to the late ISW effect from the neutrino background evolution. For the matter power spectrum, we find that massive neutrinos can suppress the amplitude of $P(k)$ for a wide range of scales, but obviously the effect on small scales is more evident than on large scales. In the lower panels, we show the cases of varying the value of $N_{\rm eff}$. We choose the examples of $\Delta N_{\rm eff}=0$, 0.5, and 1.0; the neutrino mass is fixed to be $\sum m_\nu=0.06$ eV. We find that for both CMB power spectrum $C_\ell^{TT}$ and matter power spectrum $P(k)$ the effect of dark radiation is more evident on small scales.

Note that in this example we only show some unrealistic cases with one parameter varied and others fixed; in practice in order to fit to observational data, when one parameter is changed, other parameters would also change due to some compensation and degeneracy effects. Also, in fact, the CMB alone is not very powerful for constraining neutrino mass and dark radiation, and it should be used in combination with other measurements of expansion history and growth of structure. The matter power spectrum is useful for constraining neutrino mass but in practice it has limitations due to the bias and the fact that only intermediate region can be accurately measured. Thus, better choice is to use other LSS data such as the measurements of weak lensing and redshift space distortions.

In the next section, we will use the observational data to constrain the HDE model with massive neutrinos and/or dark radiation. We will use the Planck CMB data in combination with other geometric and structural growth measurements. For the geometric measurements, we use the BAO data, SN (JLA) data, and $H_0$ measurement. For the measurements of growth of structure, we use the data of WL and RSD.

\section{Cosmological constraints}\label{sec:obs}
\label{method}

We place constraints on the models of HDE with massive neutrinos and/or extra dark radiation by using the observational data.
The conventions used in this paper are consistent with those adopted by the Planck team \cite{planck}, i.e., those used in the {\tt camb} Boltzmann code \cite{Lewis:1999bs}.
The base parameter set for the basic 7-parameter HDE model is:
$${\bf P}=\{\omega_b,~\omega_c,~100\theta_{\rm MC},~\tau,~c,~n_s,~\ln (10^{10}A_s)\},$$
where $\omega_b\equiv \Omega_b h^2$ and $\omega_c\equiv \Omega_c h^2$ are the baryon and cold dark matter densities today, respectively,
$\theta_{\rm MC}$ is the approximation used in {\tt CosmoMC}~\cite{Lewis:2002ah} to $r_s(z_\ast)/d_A(z_\ast)$ (the angular size of the sound horizon at the time of last-scattering),
$\tau$ is the Thomson scattering optical depth due to reionization,
$c$ is the parameter governing the evolution of HDE, and $n_s$ and $A_s$ are the power-law spectral index and power amplitude of the
primordial curvature perturbations, respectively.
Flat priors for the base parameters are used. Note also that the prior ranges for the base parameters are chosen to be much wider than the posterior
in order not to affect the results of parameter estimation.
We use the {\tt CosmoMC} package to infer the posterior probability distributions of parameters.

\subsection{Observational data}

Here, we describe the observational data sets used in this paper.\footnote{In our description, there are lots of acronyms. For convenience, here we give a glossary of acronyms used in the description of observational data. WMAP: Wilkinson Microwave Anisotropy Probe; 6dFGS: Six-degree-Field Galaxy Survey; SDSS: Sloan Digital Sky Survey; DR: Data Release; BOSS: Baryon Oscillation Spectroscopic Survey; JLA: Joint Light-curve Analysis; HST: Hubble Space Telescope; CFHTLenS: Canada-France-Hawaii Telescope Lensing Survey; LRG: Luminous Red Galaxy; VIPERS: VIMOS Public Extragalactic Redshift Survey.}
The data sets we use include the CMB, BAO, SN, $H_0$, WL, and RSD.

{\it The CMB data}: We use the CMB TT angular power spectrum data from the 2013 release of Planck \cite{planck}, combined with the CMB large-scale TE and EE polarization power
spectrum data form the 9-yr release of WMAP \cite{wmap9}.

{\it The BAO data}: We use the BAO measurements from the 6dFGS ($z=0.1$)~\cite{6df}, SDSS-DR7 ($z=0.35$)~\cite{sdss7}, BOSS-DR11 ($z=0.32$ and 0.57)~\cite{boss}, and
WiggleZ ($z=0.44$, 0.60, and 0.73)~\cite{wigglez} surveys. This combination of BAO data has been used widely and proven to be in good agreement with the Planck CMB data.

{\it The SN data}: We use the JLA compilation of the type Ia supernova observations containing 740 SN data \cite{JLA}.

{\it The $H_0$ measurement}: We use the direct measurement of the Hubble constant from the HST observations,
$H_0=(73.8\pm 2.4)~{\rm km}~{\rm s}^{-1}~{\rm Mpc}^{-1}$~\cite{H0}.

{\it The WL data}: We use the cosmic shear measurement of weak lensing from the CFHTLenS survey, $\sigma_8(\Omega_m/0.27)^{0.6}=0.79\pm 0.03$ \cite{shear}.
Moreover, since the CMB lensing reconstruction data directly probe the lensing power, thereby also sensitive to neutrino mass, we also use the CMB lensing power spectrum $C_\ell^{\phi\phi}$ from the Planck mission \cite{cmblensing}.

{\it The RSD data}: We use the RSD measurements of $f(z)\sigma_8(z)$ from 6dFGS ($z=0.067$) \cite{RSD6dF}, 2dFGRS ($z=0.17$) \cite{RSD2dF}, WiggleZ ($z=0.22$, 0.41, 0.60, and 0.78) \cite{RSDwigglez},
SDSS LRG DR7 ($z=0.25$ and 0.37) \cite{RSDsdss7}, BOSS CMASS DR11 ($z=0.57$) \cite{Beutler:2013yhm}, and VIPERS ($z=0.80$) \cite{RSDvipers}.

The CMB data contain both information of expansion and growth, so it is fairly important for constraining the cosmological parameters.
The BAO data probe the Hubble expansion rate and angular diameter distance at different redshifts, and have been proven to be in good agreement with the CMB data in a model-independent manner. Thus, we use the CMB+BAO as the basic data combination for constraining cosmological parameters in the HDE models. We consider the further geometric constraints by adding the SN (JLA) and $H_0$ data in the cosmological fits. Furthermore, we consider the constraints from growth of structure by adding the WL and RSD data in the fits.
We consider three models in this paper, i.e., the HDE+$\sum m_\nu$ model, the HDE+$N_{\rm eff}$ model, and the HDE+$\sum m_\nu$+$N_{\rm eff}$ model.
We will report the constraint results in the next subsection.

\subsection{Results and discussion}

\begin{figure*}
\includegraphics[width=5cm]{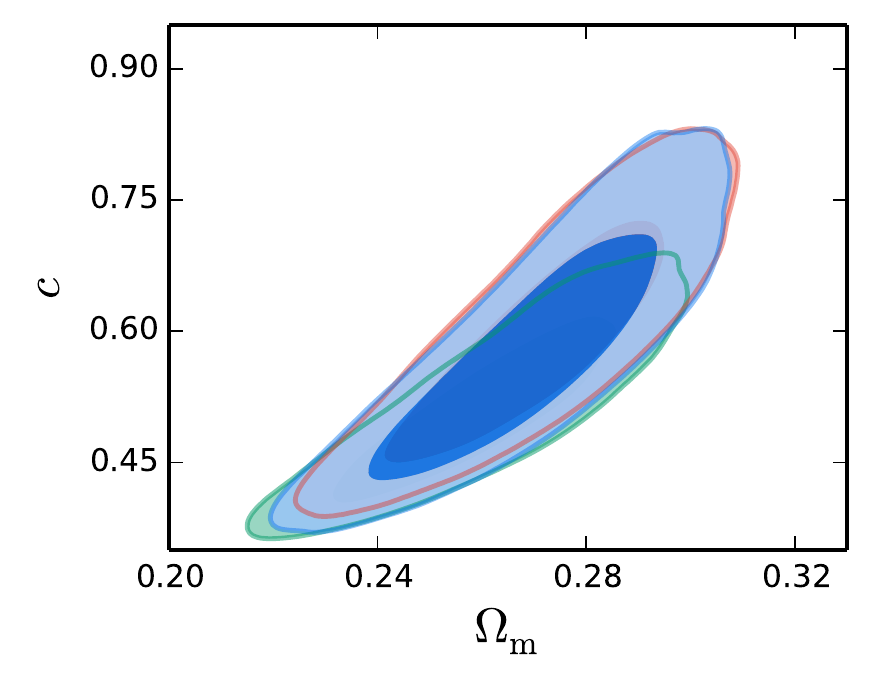}
\includegraphics[width=5cm]{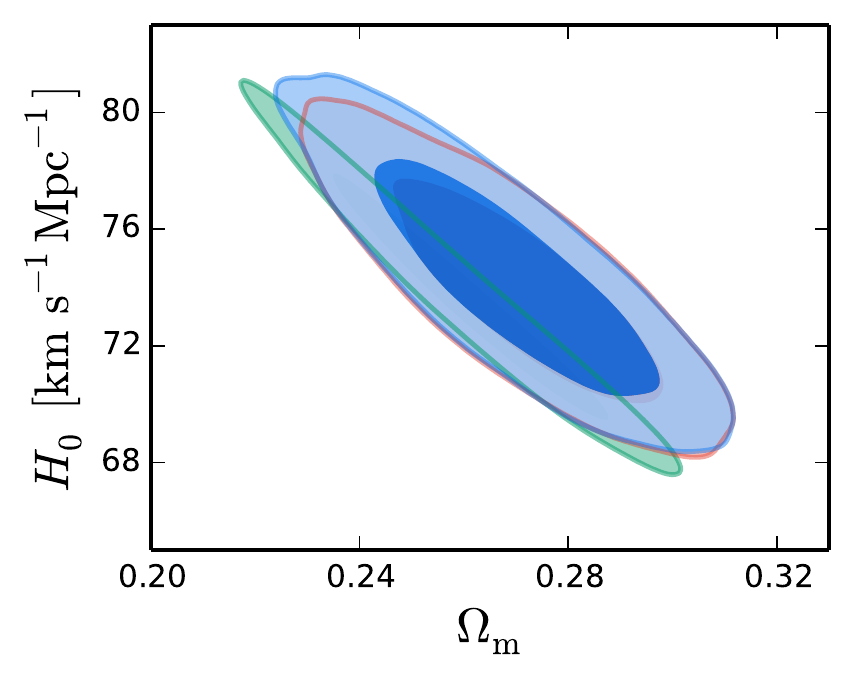}
\includegraphics[width=5cm]{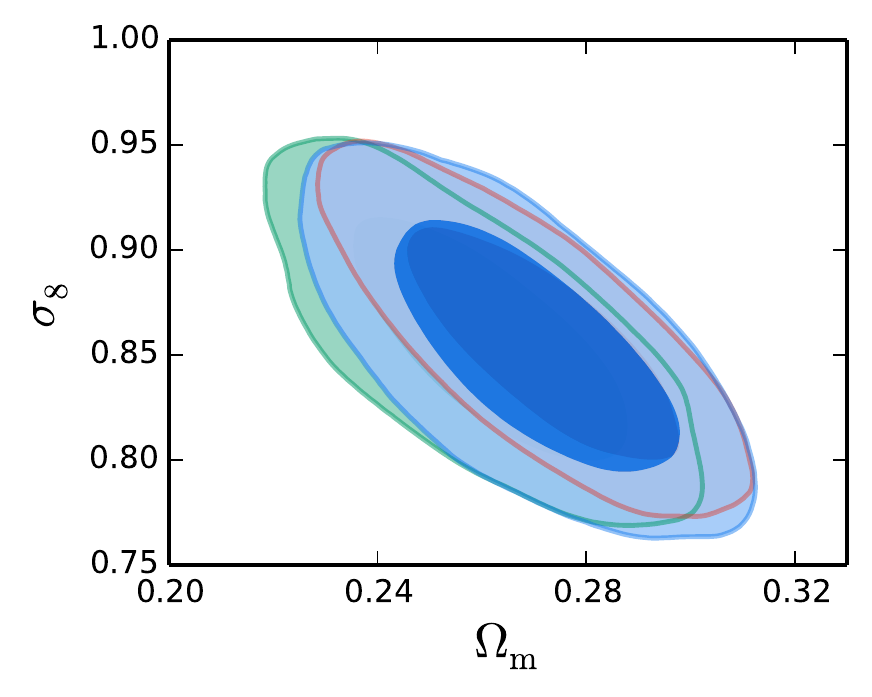}\\
\includegraphics[width=5cm]{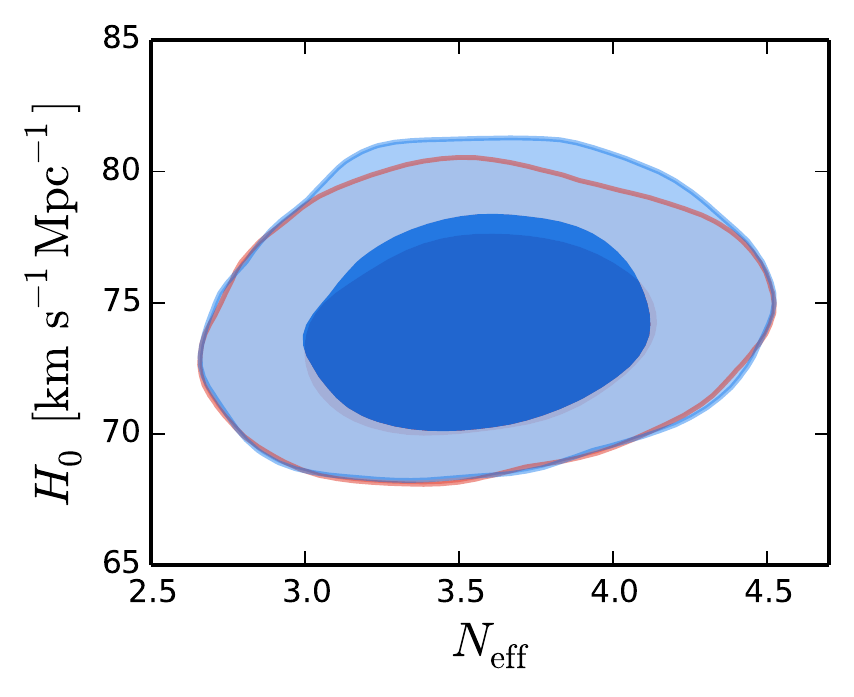}
\includegraphics[width=5.1cm]{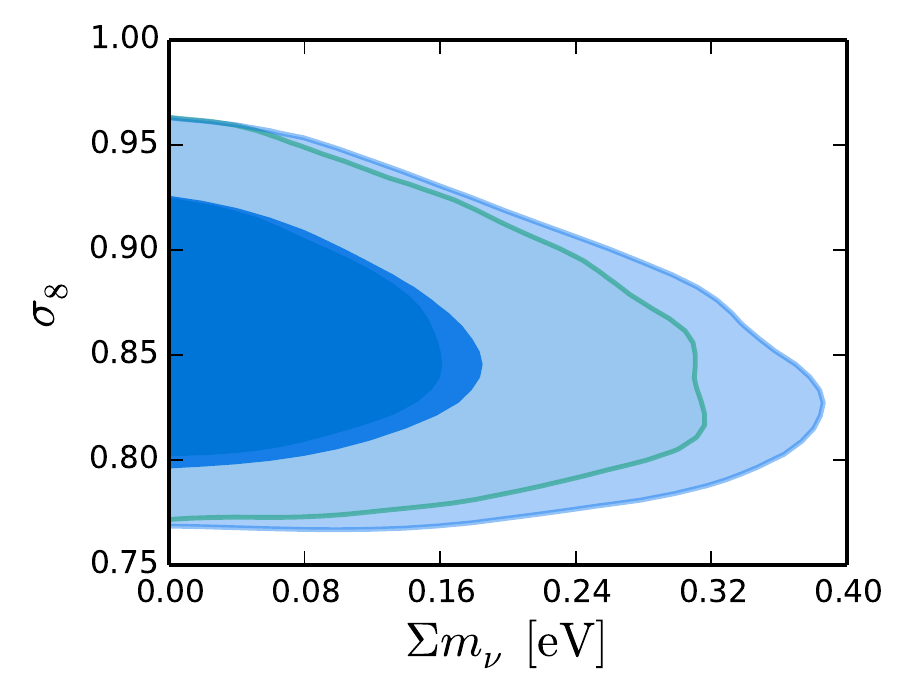}
\includegraphics[width=4.9cm]{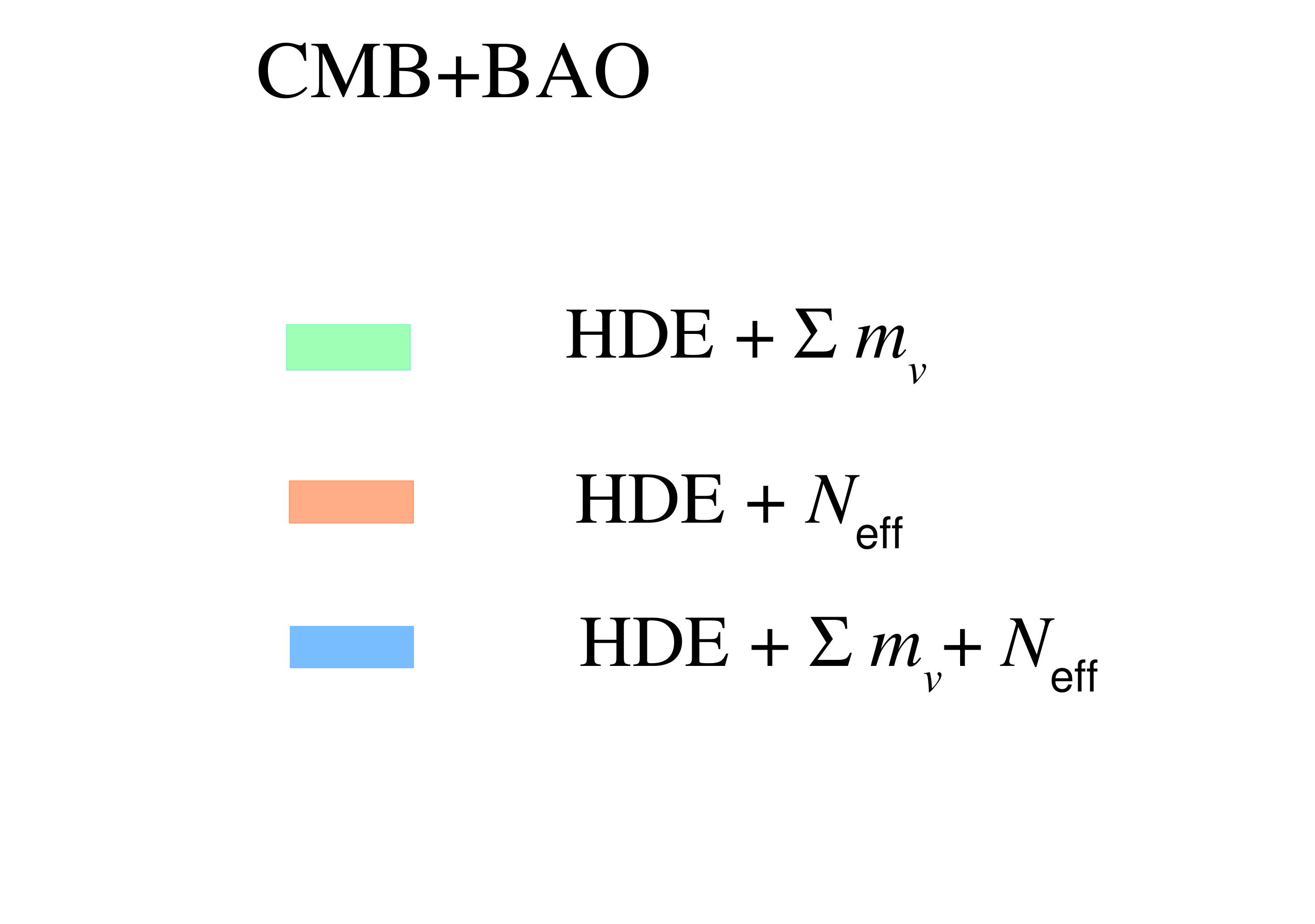}
\caption{\label{fig3}The CMB+BAO constraints on the models of HDE+$\sum m_\nu$, HDE+$N_{\rm eff}$, and HDE+$\sum m_\nu$+$N_{\rm eff}$.}
\end{figure*}

\begin{figure*}
\includegraphics[width=5.1cm]{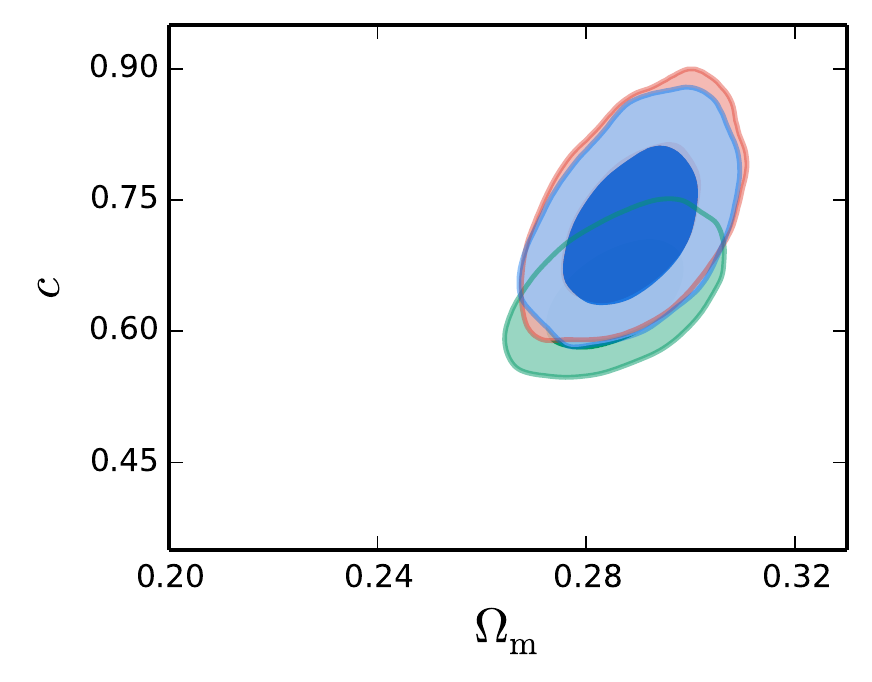}
\includegraphics[width=5cm]{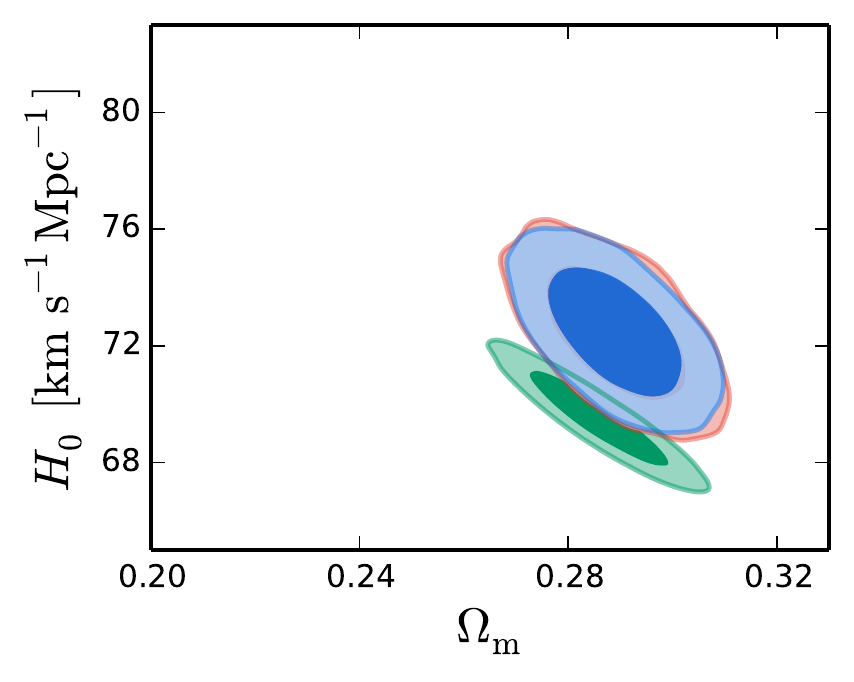}
\includegraphics[width=5.1cm]{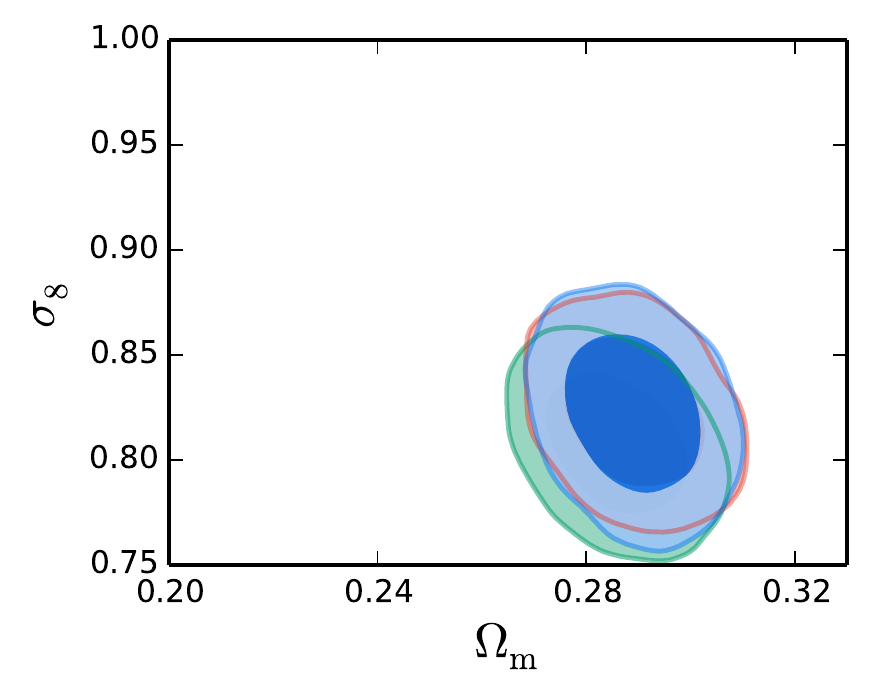}
\includegraphics[width=5cm]{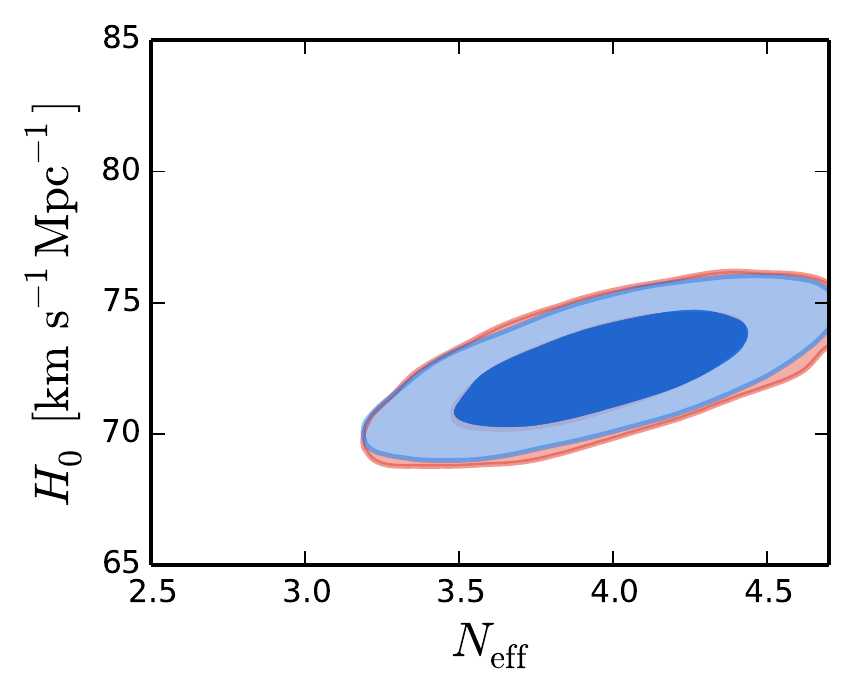}
\includegraphics[width=5.2cm]{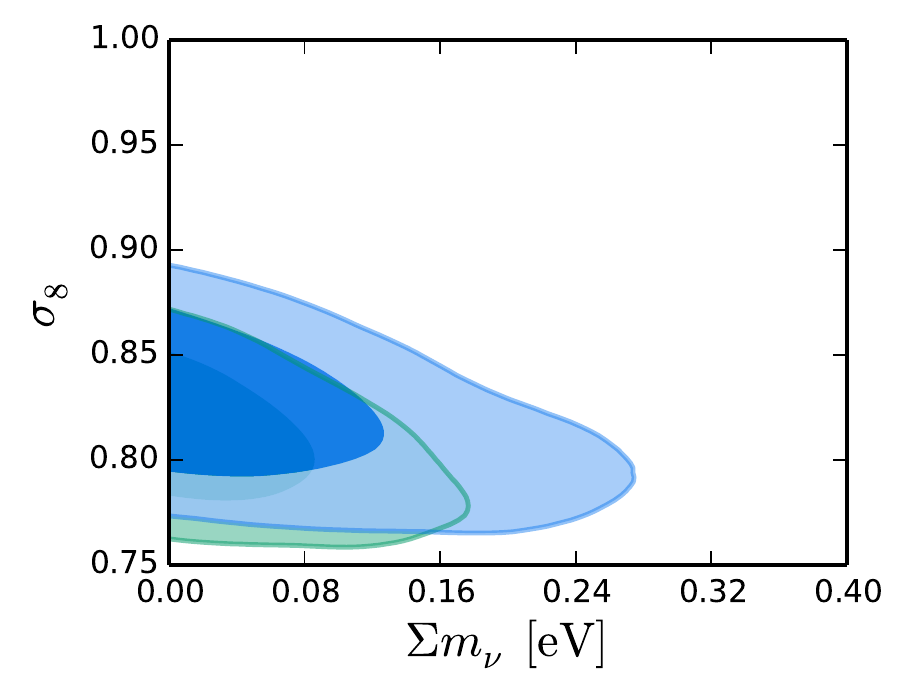}
\includegraphics[width=4.9cm]{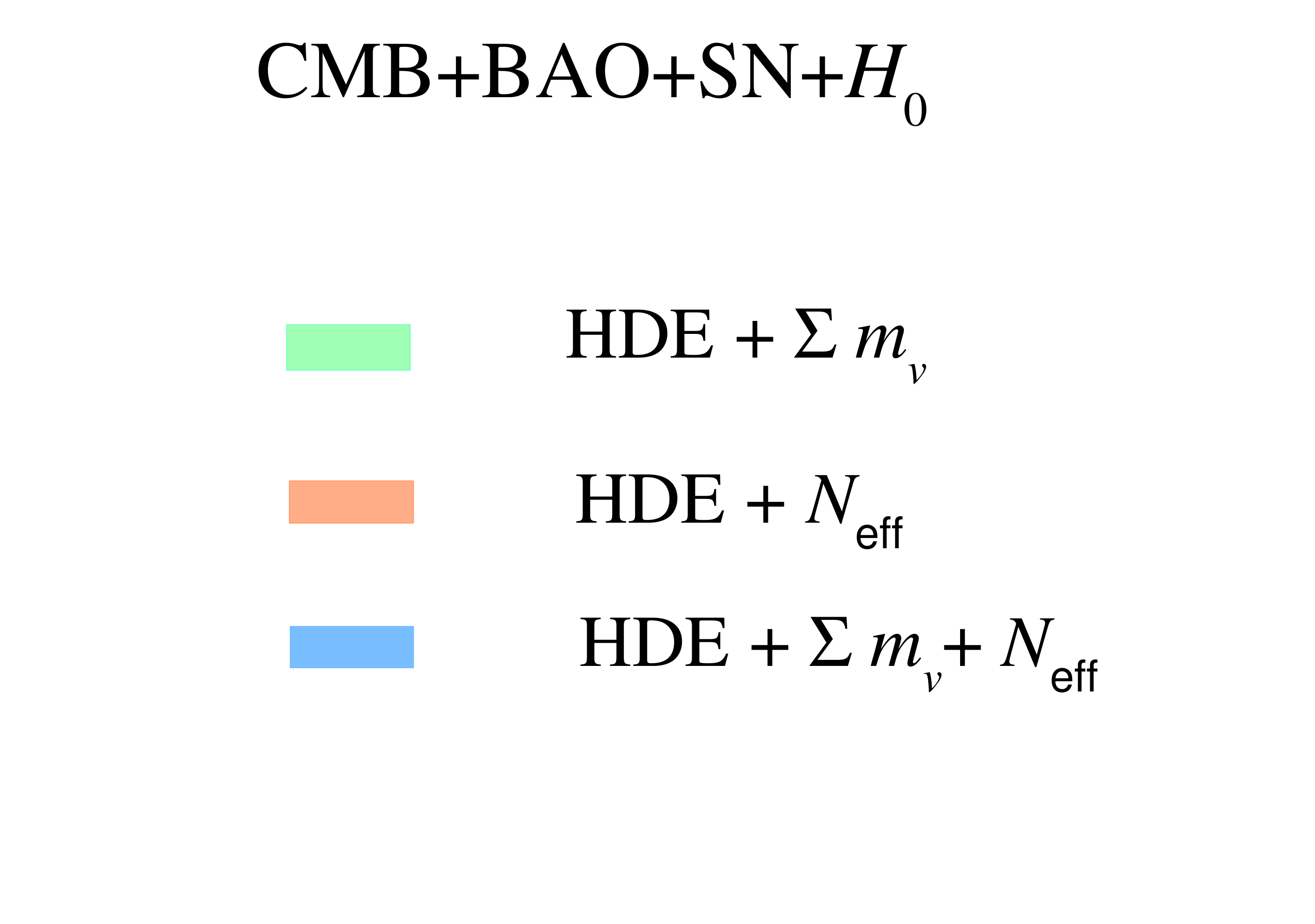}
\caption{\label{fig4}The CMB+BAO+SN+$H_0$ constraints on the models of HDE+$\sum m_\nu$, HDE+$N_{\rm eff}$, and HDE+$\sum m_\nu$+$N_{\rm eff}$.}
\end{figure*}

\begin{figure*}
\includegraphics[width=4.9cm]{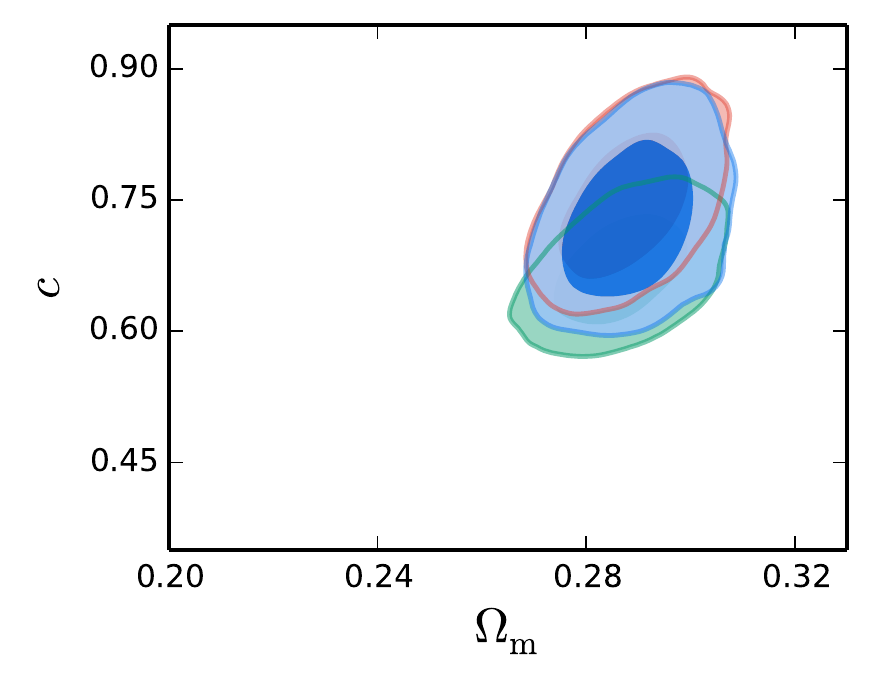}
\includegraphics[width=4.8cm]{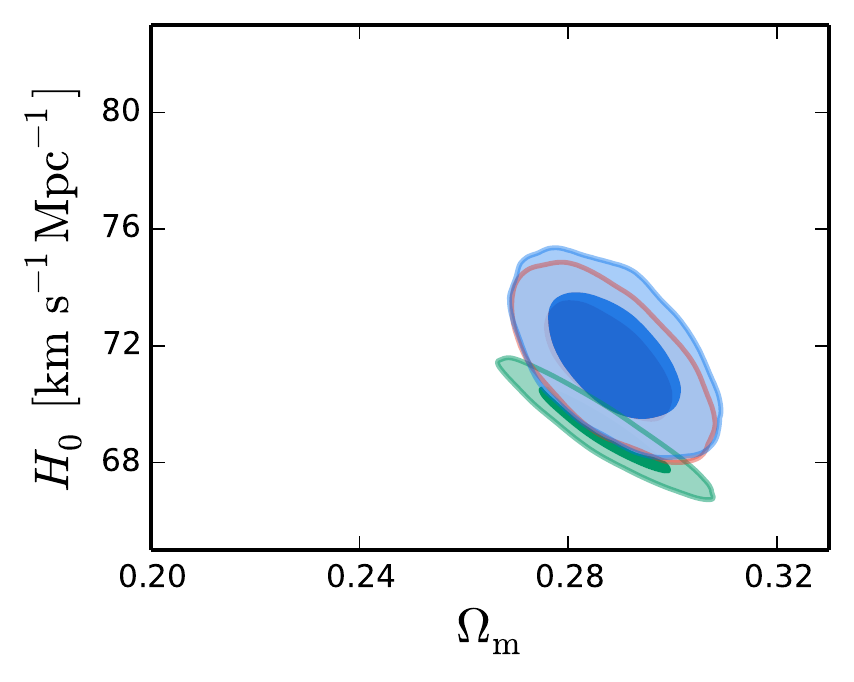}
\includegraphics[width=5cm]{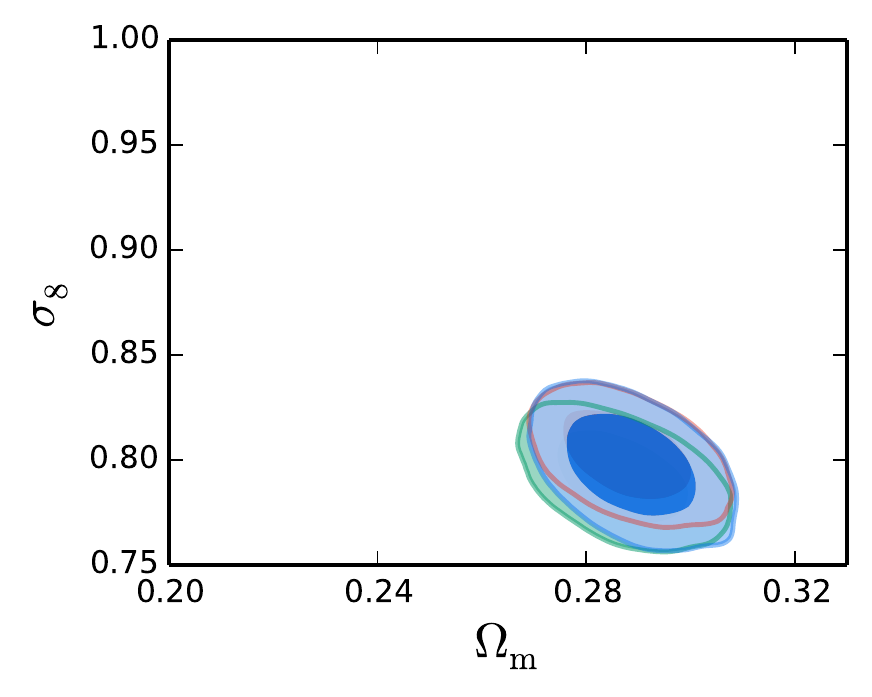}
\includegraphics[width=4.9cm]{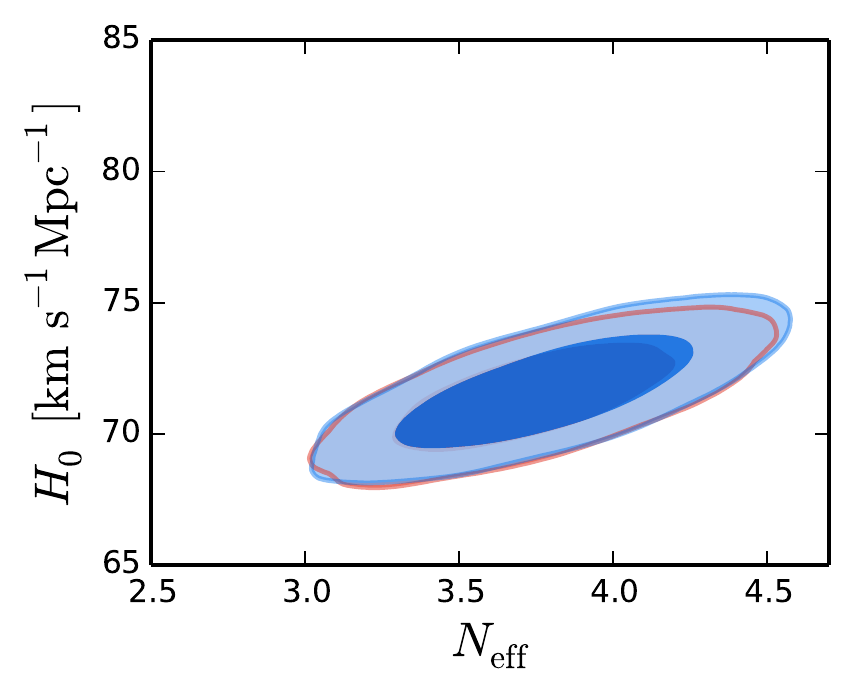}
\includegraphics[width=5.1cm]{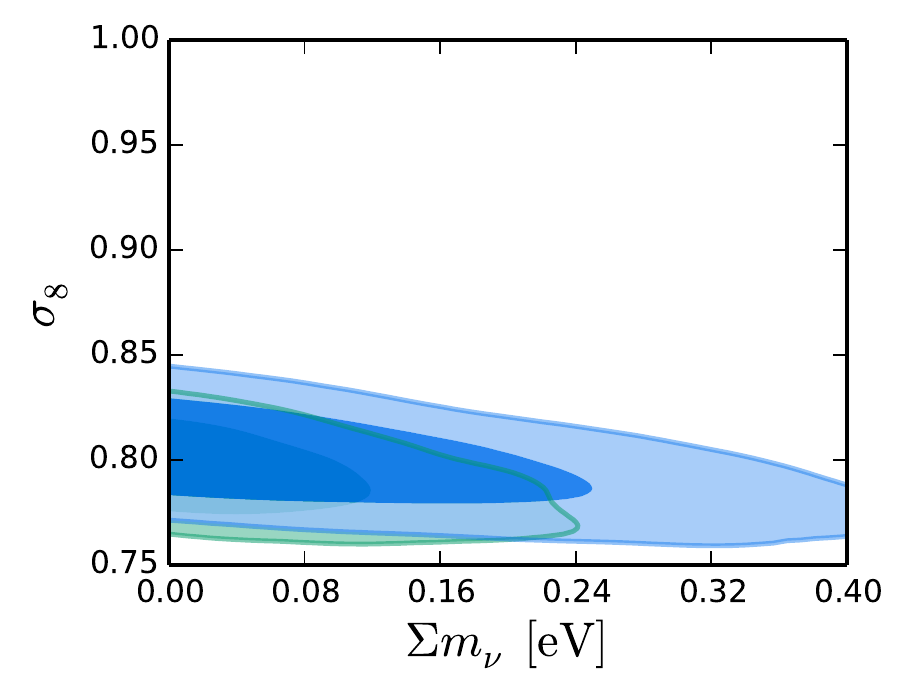}
\includegraphics[width=4.8cm]{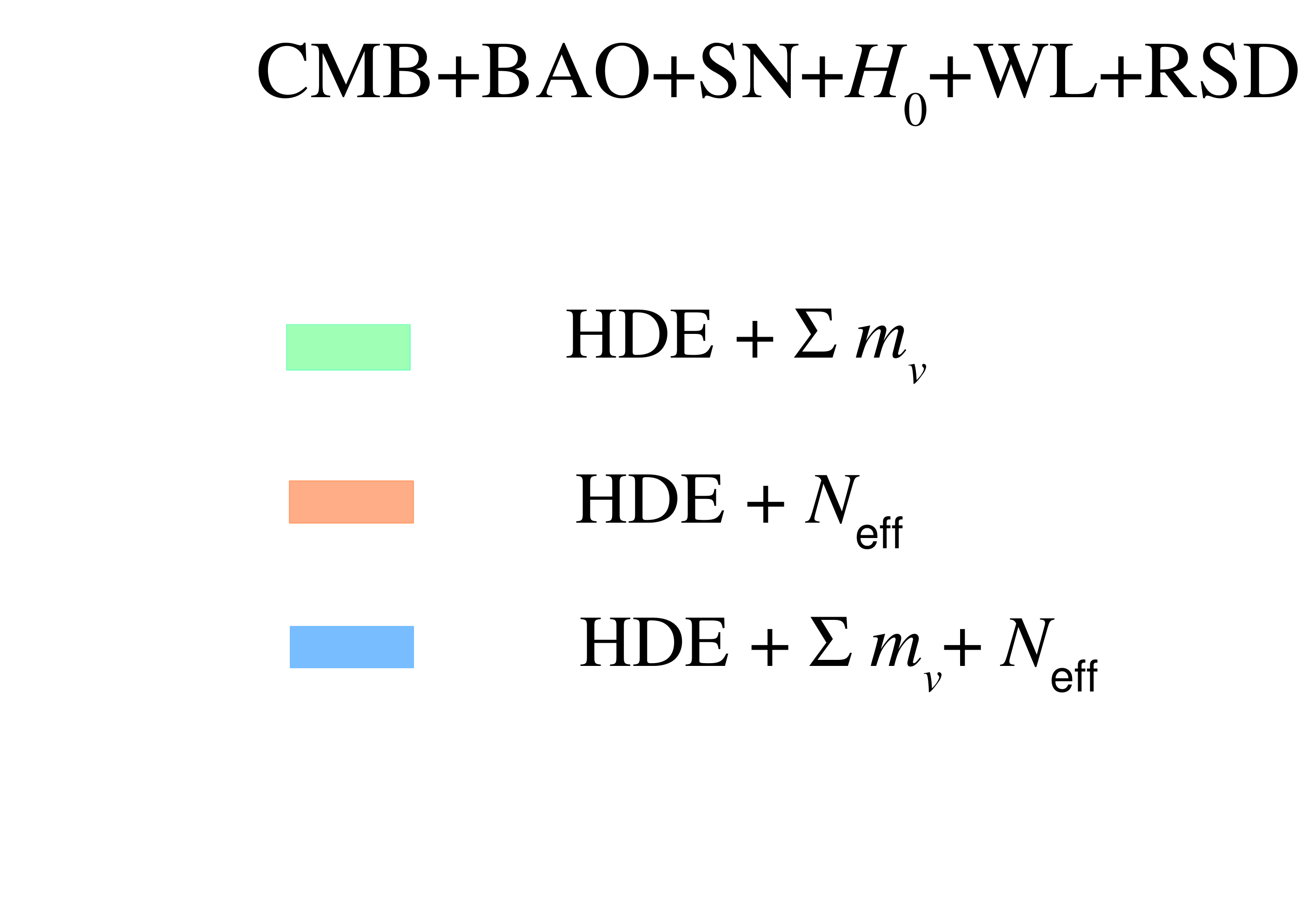}
\caption{\label{fig5}The CMB+BAO+SN+$H_0$+WL+RSD constraints on the models of HDE+$\sum m_\nu$, HDE+$N_{\rm eff}$, and HDE+$\sum m_\nu$+$N_{\rm eff}$. }
\centering
\end{figure*}

\begin{table*}\small
\begin{tabular}{ccccccccc}
\hline &\multicolumn{2}{c}{ HDE+$\sum m_{\nu}$}&&\multicolumn{2}{c}{HDE+$N_{\rm{eff}}$}&&\multicolumn{2}{c}{HDE+$\sum m_{\nu}+N_{\rm{eff}}$}\\
           \cline{2-3}\cline{5-6}\cline{8-9}
Parameter  & Best fit & $68\% $ limits&& Best fit & $68\% $ limits &&  Best fit & $68\% $ limits \\
\hline
$\Omega_bh^2$&$0.02196$&$0.02213^{+0.00026}_{-0.00025}$&&$0.02246$&$0.02246\pm0.00035$&&$0 .02225$&$0.02245^{+0.00035}_{-0.00038}$\\
$\Omega_ch^2$&$0.1197$&$0.1178^{+0.0022}_{-0.0020}$&&$0.1228$&$0.1249^{+0.0051}_{-0.0058}$&&$0 .1265$&$0.1249^{+0.0053}_{-0.0052}$\\
$100\theta_{\rm MC}$&$1.04118$&$1.0415^{+0.00058}_{-0.00059}$&&$1.04095$&$1.04083\pm0.00074$&&$1 .0405$&$1.04079^{+0.00073}_{-0.00072}$\\
$\tau$&$0.083$&$0.094^{+0.013}_{-0.014}$&&$0.104$&$0.099^{+0.014}_{-0.016}$&&$0 .093$&$0.099^{+0.014}_{-0.016}$\\
$n_s$&$0.9619$&$0.9636\pm0.0064$&&$0.983$&$0 .982\pm0.014$&&$0.975$&$0.981^{+0.014}_{-0.015}$\\
$c$&$0.462$&$0.515^{+0.060}_{-0.075}$&&$0.571$&$0.595^{+0.077}_{-0.103}$&&$0 .524$&$0 .578^{+0.076}_{-0.110}$\\
$\sum m_\nu$&$0.033$&$<0.250$&&...&...&&$0 .007$&$<0.299$\\
$N_{\rm{eff}}$&...&...&&$3.45$&$3.56\pm0.36$&&$3 .54$&$3 .57\pm0.36$\\
$\Omega_m$&$0.245$&$0.261^{+0.018}_{-0.016}$&&$0.265$&$0.271^{+0.018}_{-0.016}$&&$0 .255$&$0.269\pm0.017$\\
$\sigma_8$&$0.897$&$0.857\pm0.036$&&$0.863$&$0.857^{+0.033}_{-0.039}$&&$0 .899$&$0 .853^{+0.037}_{-0.040}$\\
$H_0$&$76.1$&$73.6^{+2.4}_{-3.1}$&&$74.3$&$74.0^{+2.2}_{-2.7}$&&$76 .4$&$74.4^{+2.4}_{-3.1}$\\
\hline
$-\ln\mathcal{L}_{\rm{max}}$ &\multicolumn{2}{c}{4906.71} & & \multicolumn{2}{c}{4906.56} & & \multicolumn{2}{c}{4906.47} \\
\hline
\end{tabular}
\caption{\label{tab1}Fitting results from the CMB+BAO data. We quote the $\pm1\sigma$ errors, but for the neutrino mass $\sum m_\nu$, we quote the 95\% CL upper limits.
Note that $\sum m_\nu$ is in unit of eV and $H_0$ is in unit of km~s$^{-1}$~Mpc$^{-1}$. }
\end{table*}

\begin{table*}\small
\begin{tabular}{ccccccccc}
\hline &\multicolumn{2}{c}{ HDE+$\sum m_{\nu}$}&&\multicolumn{2}{c}{HDE+$N_{\rm{eff}}$}&&\multicolumn{2}{c}{HDE+$\sum m_{\nu}+N_{\rm{eff}}$}\\
           \cline{2-3}\cline{5-6}\cline{8-9}
Parameter  & Best fit & $68\% $ limits&& Best fit & $68\% $ limits &&  Best fit & $68\% $ limits \\ \hline
$\Omega_bh^2$&$0.02228$&$0.02234\pm0.00025$&&$0.02292$&$0.02276^{+0.00033}_{-0.00038}$&&$0.02271$&$0.02283\pm0.0003$\\
$\Omega_ch^2$&$0.116$&$0.1151\pm0.0017$&&$0.1303$&$0.1267^{+0.0079}_{-0.0072}$&&$0.1256$&$0.1284^{+0.0047}_{-0.0048}$\\
$100\theta_{\rm MC}$&$1.04181$&$1.04193^{+0.00055}_{-0.00056}$&&$1.04011$&$1.0408^{+0.00067}_{-0.00085}$&&$1.04059$&$1.04062^{+0.00067}_{-0.00073}$\\
$\tau$&$0.105$&$0.099^{+0.013}_{-0.015}$&&$0.114$&$0.106^{+0.014}_{-0.019}$&&$0.107$&$0.107^{+0.015}_{-0.017}$\\
$n_s$&$0.9677$&$0.9705\pm0.0057$&&$1.008$&$0.995\pm0.015$&&$0.992$&$0.999\pm0.011$\\
$c$&$0.649$&$0.643^{+0.038}_{-0.044}$&&$0.757$&$0.715^{+0.053}_{-0.069}$&&$0.713$&$0.722^{+0.052}_{-0.064}$\\
$\sum m_\nu$&$0.0009$&$<0.133$&&...&...&&$0.009$&$<0.207$\\
$N_{\rm{eff}}$&...&...&&$4.18$&$3.84^{+0.43}_{-0.42}$&&$3.77$&$3.97^{+0.30}_{-0.31}$\\
$\Omega_m$&$0.2895$&$0.2859\pm0.0084$&&$0.2882$&$0.289\pm0.0083$&&$0.2871$&$0.2891\pm0.0081$\\
$\sigma_8$&$0.823$&$0.808\pm0.021$&&$0.823$&$0.821^{+0.024}_{-0.026}$&&$0.830$&$0.822^{+0.026}_{-0.023}$\\
$H_0$&$69.1$&$69.5^{+1.2}_{-1.1}$&&$73.0$&$72.1\pm1.8$&&$71.9$&$72.5\pm1.4$\\
\hline
$-\ln\mathcal{L}_{\rm{max}}$ &\multicolumn{2}{c}{5259.69} & & \multicolumn{2}{c}{5256.35} & & \multicolumn{2}{c}{5256.23} \\
\hline
\end{tabular}
\caption{\label{tab2}Fitting results from the CMB+BAO+SN+$H_0$ data. We quote the $\pm1\sigma$ errors, but for the neutrino mass $\sum m_\nu$, we quote the 95\% CL upper limits.
Note that $\sum m_\nu$ is in unit of eV and $H_0$ is in unit of km~s$^{-1}$~Mpc$^{-1}$.}
\end{table*}

\begin{table*}\small
\centering
\begin{tabular}{ccccccccc}
\hline &\multicolumn{2}{c}{ HDE+$\sum m_{\nu}$}&&\multicolumn{2}{c}{HDE+$N_{\rm{eff}}$}&&\multicolumn{2}{c}{HDE+$\sum m_{\nu}+N_{\rm{eff}}$}\\
           \cline{2-3}\cline{5-6}\cline{8-9}
Parameter  & Best fit & $68\% $ limits&& Best fit & $68\% $ limits &&  Best fit & $68\% $ limits \\
\hline
 $\Omega_bh^2$&$0.02225$&$0.02249\pm0.00024$&&$0.02276$&$0.02292\pm0.0003$&&$0.02297$&$0.02292\pm0.0003$\\
 $\Omega_ch^2$&$0.1145$&$0.1137\pm0.0013$&&$0.122$&$0.1234^{+0.0041}_{-0.0045}$&&$0.1236$&$0.1239\pm0.0043$\\
 $100\theta_{\rm MC}$&$1.04165$&$1.04203\pm0.00056$&&$1.04076$&$1.04098^{+0.00066}_{-0.00065}$&&$1.04119$&$1.04095^{+0.00067}_{-0.00068}$\\
$\tau$&$0.092$&$0.102^{+0.013}_{-0.014}$&&$0.115$&$0.107^{+0.014}_{-0.016}$&&$0 .109$&$0 .111^{+0.015}_{-0.017}$\\
$n_s$&$0.9700$&$0.9735^{+0.0056}_{-0.0055}$&&$0.992$&$0.996\pm0.011$&&$0.996$&$0.998\pm0.011$\\
$c$&$0.658$&$0.670\pm0.040$&&$0.728$&$0.747^{+0.048}_{-0.060}$&&$0 .769$&$0 .728^{+0.051}_{-0.066}$\\
$\sum m_\nu$&$0.027$&$<0.186$&&...&...&&$0.003$&$<0.343$\\
$N_{\rm{eff}}$&...&...&&$3.68$&$3.75^{+0.28}_{-0.32}$&&$3 .78$&$3 .79\pm0.30$\\
$\Omega_m$&$0.2860$&$0.2869\pm0.0080$&&$0.2846$&$0.2879^{+0.0075}_{-0.0085}$&&$0 .2852$&$0 .2887^{+0.0078}_{-0.0085}$\\
$\sigma_8$&$0.800$&$0.792\pm0.014$&&$0.809$&$0.802^{+0.013}_{-0.014}$&&$0 .810$&$0 .797^{+0.017}_{-0.015}$\\
$H_0$&$69.21$&$69.10^{+0.94}_{-0.95}$&&$71.5$&$71.4\pm1.3$&&$71 .7$&$71 .7^{+1.4}_{-1.5}$\\
\hline
$-\ln\mathcal{L}_{\rm{max}}$ &\multicolumn{2}{c}{5270.86} & & \multicolumn{2}{c}{5268.35} & & \multicolumn{2}{c}{5268.29} \\
\hline
\end{tabular}
\caption{\label{tab3}Fitting results from the CMB+BAO+SN+$H_0$+WL+RSD data. We quote the $\pm1\sigma$ errors, but for the neutrino mass $\sum m_\nu$, we quote the 95\% CL upper limits.
Note that $\sum m_\nu$ is in unit of eV and $H_0$ is in unit of km~s$^{-1}$~Mpc$^{-1}$.}
\end{table*}

We report the results for the HDE+$\sum m_\nu$ model (Model I), the HDE+$N_{\rm eff}$ model (Model II), and the HDE+$\sum m_\nu$+$N_{\rm eff}$ model (Model III) fitting to the three data combinations, i.e., the CMB+BAO (D1) combination, the CMB+BAO+SN+$H_0$ (D2) combination, and the CMB+BAO+SN+$H_0$+WL+RSD (D3) combination. For convenience, in the text of this subsection, we use Model I, Model II, and Model III to denote the three models, and use D1, D2, and D3 to denote the three data combinations.

The constraint results are shown in Figs.~\ref{fig3}--\ref{fig5} and Tables \ref{tab1}--\ref{tab3}. Fig. \ref{fig3} and Table \ref{tab1} are responsible for the D1 constraints, Fig. \ref{fig4} and Table \ref{tab2} are for the D2 constraints, and Fig. \ref{fig5} and Table \ref{tab3} are for the D3 constraints. In each figure, we show the two-dimensional posterior distribution contours (68\% and 95\% CL) in the $\Omega_m$--$c$, $\Omega_m$--$H_0$, $\Omega_m$--$\sigma_8$, $N_{\rm eff}$--$H_0$, and $\sum m_\nu$--$\sigma_8$ planes. Contours in green, orange, and blue stand for Model I, Model II, and Model III, respectively. One can easily compare different models under the same data combination in each individual figure, and it is also convenient to make a comparison for the constraint results of the same model with different data combinations by comparing the corresponding panels in different figures.

First, we discuss the constraint results from the D1 combination. From Fig. \ref{fig3}, one can see that there are strong parameter degeneracies in all the parameter planes. This is mainly due to the fact that using only CMB and BAO data cannot accurately measure these cosmological parameters in a dynamical dark energy model with neutrinos/dark radiation. The Planck data accurately measure seven acoustic peaks, and so the observed angular size of acoustic scale $\theta_\ast=r_s/D_A$ is determined to a very high precision (better than 0.1\% precision at 1$\sigma$), implying tight constraints on some combinations of the cosmological parameters that determine $r_s$ and $D_A$. The degeneracies in the $\Omega_m$--$c$ and $\Omega_m$--$H_0$ planes can be viewed as mainly coming from the fact that the parameter combinations fitting to the Planck data must be constrained to be close to a surface of constant $\theta_\ast$. The BAO data have helped reduce the geometrical degeneracies to some extent, but not enough. Additional data are needed to further break the degeneracies, which will be seen in the following discussions. 

We find that for all the three models, the degeneracy directions are the same. But we find that in the $\Omega_m$--$c$ and $\Omega_m$--$H_0$ planes the contours of Model I (green) are somewhat lower, comparing to those of Model II (orange). Both neutrino mass and extra radiation could affect the shape of the CMB power spectra to some degree, but their effects are subtly different, as mentioned in the previous section. The effects on the background cosmology can be compensated by changes in, say, $H_0$ (and $c$), to ensure the same observed acoustic peak scale $\theta_\ast$. Due to the subtle difference in the compensation effects, the contours of Model I and Model II exhibit some differences. Comparing the fittings of the two models, we find that Model II is better than Model I by $\Delta \chi^2=0.3$. Though Model III contains two extra parameters, $\sum m_\nu$ and $N_{\rm eff}$, it is only slightly better than Model II in the fit by $\Delta \chi^2=0.18$, implying that $N_{\rm eff}$ dominates over $\sum m_\nu$ in the fit and explaining why the contours of Model II and Model III are similar. 

In the case of D1 constraints, we find that $c=0.515^{+0.060}_{-0.075}$ for Model I, $c=0.595^{+0.077}_{-0.103}$ for Model II, and $c=0.578^{+0.076}_{-0.110}$ for Model III. We thus find that lower values of $c$ are derived once we only use the CMB and BAO data, consistent with the results of Ref. \cite{hde20}. We can constrain $N_{\rm eff}$ well but can only obtain an upper limit for $\sum m_\nu$. We find that $\sum m_\nu<0.250$ eV (95\% CL) for Model I and $\sum m_\nu<0.299$ eV (95\% CL) for Model III;  $N_{\rm eff}=3.56\pm 0.36$ for Model II and $N_{\rm eff}=3.57\pm 0.36$ for Model III.

Second, we discuss the constraint results from the D2 combination. From Fig. \ref{fig4}, we find that all the contours are shrunk evidently, indicating that the constraint results are greatly tightened by adding more data of expansion history, especially the SN JLA data. It is clear to see that in this case the degeneracies are broken substantially due to the use of the SN data that have outstanding power for probing the properties of dark energy. In this fit, we find that Model II is much better than Model I by $\Delta\chi^2=6.68$, but Model III is only slightly better than Model II by $\Delta\chi^2=0.24$. This feature strongly indicates that in the fit $N_{\rm eff}$ dominates over $\sum m_\nu$, as demonstrated in the previous case.

Under the constraints of D2 combination, for the HDE parameter, we find that $c=0.643^{+0.038}_{-0.044}$ for Model I, $c=0.715^{+0.053}_{-0.069}$ for Model II, and $c=0.722^{+0.052}_{-0.064}$ for Model III. So one can see that the values of $c$ are enhanced once more geometric measurements data are added. For the parameters of massive neutrinos and dark radiation, we find that $\sum m_\nu<0.133$ eV (95\% CL) for Model I and $\sum m_\nu<0.207$ eV (95\% CL) for Model III;  $N_{\rm eff}=3.84^{+0.43}_{-0.42}$ for Model II and $N_{\rm eff}=3.97^{+0.30}_{-0.31}$ for Model III. So, in this case, $\Delta N_{\rm eff}>0$ is constrained to be at about the 3$\sigma$ level.

Finally, we discuss the constraint results from the D3 combination. From Fig. \ref{fig5}, we find that the contours are further slightly shrunk by adding the measurements data of growth of structure, WL and RSD. Comparing Figs. \ref{fig3}--\ref{fig5}, we notice that the inclusion of additional background data (SN+$H_0$) gives a visible improvement in the constraints, while the improvement is not evident when the growth data (WL+RSD) are further added. This is because we are considering a smooth dark energy model where the perturbations are suppressed on small scales. For the density perturbations in the HDE model, recall the examples shown in Fig. \ref{fig1}. Thus, the constraints from the structure growth probes are weaker. But we indeed find that in this case the fit values of $\sigma_8$ are further suppressed due to the consideration of the growth data. In this fit, Model II is better than Model I by $\Delta\chi^2=5.02$, and Model III is better than Model II by $\Delta\chi^2=0.12$.

For this case, we find that $c=0.670\pm0.040$ for Model I, $c=0.747^{+0.048}_{-0.060}$ for Model II, and $c=0.728^{+0.051}_{-0.066}$ for Model III; we find that $\sum m_\nu<0.186$ eV (95\% CL) for Model I and $\sum m_\nu<0.343$ eV (95\% CL) for Model III;  $N_{\rm eff}=3.75^{+0.28}_{-0.32}$ for Model II and $N_{\rm eff}=3.79\pm0.30$ for Model III. In this case, $\Delta N_{\rm eff}>0$ is still constrained to be at more than 2$\sigma$ level. Since the LSS data (WL and RSD) prefer lower $\sigma_8$, this leads to the upper limit on $\sum m_\nu$ (comparing to the second case) being enhanced slightly.

\section{Conclusion}\label{sec:concl}

In this paper, we investigated the models of HDE with massive neutrinos and/or dark radiation in detail. We calculated the background and perturbation evolutions in the HDE model. When the parameter $c$ of HDE is less than 1, the EOS parameter $w$ will evolve across the phantom divide $w=-1$, which usually leads to the perturbation divergence of dark energy. In order to overcome this instability difficulty, we employed the PPF approach in the perturbation calculations. We derived the evolutions of density perturbations of various components and metric fluctuations in the HDE model. We further discussed the impacts of massive neutrino and dark radiation on the CMB anisotropy power spectrum $C_\ell^{TT}$ and the matter power spectrum $P(k)$ in the HDE scenario.

Furthermore, we constrained the models of HDE + $\sum m_\nu$, HDE + $N_{\rm eff}$, and HDE + $\sum m_\nu$ + $N_{\rm eff}$ (denoted as Model I, Model II, and Model III, respectively) by using the data combinations of CMB+BAO, CMB+BAO+SN+$H_0$, and CMB+BAO+SN+$H_0$+WL+RSD (denoted as D1, D2, and D3 combinations, respectively). With the latest measurements of expansion history and growth of structure, we found that the HDE models can be tightly constrained. 

When we considered the D1 combination, we found that there are strong degeneracies among the cosmological parameters. Since the Planck data accurately determine the acoustic scale $\theta_\ast$ to a very high precision, some parameter combinations that determine $r_s$ and $D_A$ are tightly constrained, but this also leads to some degeneracies due to the fact that the parameter combinations must be constrained to be close to a surface of constant $\theta_\ast$. Though with the help of BAO data, some degeneracies have been reduced to some extent, this is not enough, and more additional data are needed to further break the degeneracies. Thus we further considered the additional background expansion data, SN (JLA) and $H_0$. Under the constraints of D2, we found that the degeneracies are broken well. This indicates that for constraining a dynamical dark energy model (with neutrinos/dark radiation), the background data, especially the SN data, play a crucial role. Furthermore, we also considered the structure growth data, WL and RSD. We found that in the D3 fit the constraints are only slightly improved comparing to the D2 case, showing that comparing with the expansion probes, the constraints from growth probes are weaker. This is because we are considering a smooth dark energy model where the perturbations are suppressed on small scales. In addition, in all the cases, we found that in our cosmological fits Model II is much better than Model I, while Model III is only slightly better than Model II (though the former with one more parameter). This implies that $N_{\rm eff}$ dominates over $\sum m_\nu$ in the fits.

Using the CMB+BAO+SN+$H_0$+WL+RSD data, we obtained the results: $c=0.670\pm0.040$ and $\sum m_\nu<0.186$ eV (95\% CL) for the HDE+$\sum m_\nu$ model, $c=0.747^{+0.048}_{-0.060}$ and $N_{\rm eff}=3.75^{+0.28}_{-0.32}$ for the HDE+$N_{\rm eff}$ model, and $c=0.728^{+0.051}_{-0.066}$, $\sum m_\nu<0.343$ eV (95\% CL), and $N_{\rm eff}=3.79\pm0.30$ for the HDE+$\sum m_\nu$+$N_{\rm eff}$ model.

\acknowledgments

This work was supported by the National Natural Science Foundation of
China under Grant No.~11175042, the Provincial Department of Education of
Liaoning under Grant No.~L2012087, and the Fundamental Research Funds for the
Central Universities under Grants No.~N140505002, No.~N140506002, and No.~N140504007.



\end{document}